\documentclass{aa}  
\usepackage{graphicx}
\usepackage{txfonts}
\usepackage[dvipsnames]{xcolor}
\usepackage[colorlinks=true,citecolor=blue,linkcolor=blue,urlcolor=cyan]{hyperref}
\usepackage{pdflscape}
\newcommand{\nodata}{}
\renewcommand\labelitemii{$\m@th\bullet$}
\begin{document} 

\title{Corona-Australis DANCe}
\subtitle{I. Revisiting the census of stars with Gaia-DR2 data \thanks{Tables~\ref{tab_members}, \ref{tab_prob} and \ref{tab_isochrone} are only available in electronic form
at the CDS via anonymous ftp to cdsarc.u-strasbg.fr (130.79.128.5) or via http://cdsweb.u-strasbg.fr/cgi-bin/qcat?J/A+A/}}

\author{
P.A.B.~Galli \inst{1}
\and
H.~Bouy \inst{1}
\and
J.~Olivares\inst{1}
\and 
N.~Miret-Roig\inst{1}
\and
L.M.~Sarro\inst{2}
\and
D.~Barrado\inst{3}
\and
A.~Berihuete\inst{4}
\and 
W. Brandner\inst{5}
}

\institute{
Laboratoire d’Astrophysique de Bordeaux, Univ. Bordeaux, CNRS, B18N, allée Geoffroy Saint-Hillaire, F-33615 Pessac, France\\
\email{phillip.galli@u-bordeaux.fr}
\and
Depto. de Inteligencia Artificial, UNED, Juan del Rosal, 16, 28040 Madrid, Spain
\and
Centro de Astrobiolog\'ia, Depto. de Astrof\'isica, INTA-CSIC, ESAC Campus, Camino Bajo del Castillo s/n, 28692 Villanueva de la Ca\~nada, Madrid, Spain
\and
Dept. Statistics and Operations Research, University of C\'adiz, Campus Universitario R\'io San Pedro s/n, 11510 Puerto Real, \\C\'adiz, Spain
\and
Max Planck Institute for Astronomy, Königstuhl 17, 69117, Heidelberg, Germany.
}

\date{Received September 15, 1996; accepted March 16, 1997}

 \abstract
{Corona-Australis is one of the nearest regions to the Sun with recent and ongoing star formation, but the current picture of its stellar (and substellar) content is not complete yet.}
{We take advantage of the second data release of the \textit{Gaia} space mission to revisit the stellar census and search for additional members of the young stellar association in Corona-Australis.}
{We applied a probabilistic method to infer membership probabilities based on a multidimensional astrometric and photometric data set over a field of 128~deg$^{2}$  around the dark clouds of the region.}
{We identify 313 high-probability candidate members to the Corona-Australis association, 262 of which had never been reported as members before. Our sample of members covers the magnitude range between $G\gtrsim5$~mag and $G\lesssim20$~mag, and it reveals the existence of two kinematically and spatially distinct subgroups. There is a distributed `off-cloud' population of stars located in the north of the dark clouds that is twice as numerous as the historically known `on-cloud' population that is concentrated around the densest cores. By comparing the location of the stars in the HR-diagram with evolutionary models, we show that these two populations are younger than 10~Myr. Based on their infrared excess emission, we identify 28~Class~II and 215~Class~III stars among the sources with available infrared photometry, and we conclude that the frequency of Class~II stars (i.e. `disc-bearing' stars) in the on-cloud region is twice as large as compared to the off-cloud population. The distance derived for the Corona-Australis region based on this updated census is $d=149.4^{+0.4}_{-0.4}$~pc, which exceeds previous estimates by about 20~pc.}
{In this paper we provide the most complete census of stars in Corona-Australis available to date that can be confirmed with \textit{Gaia} data. Furthermore, we report on the discovery of an extended and more evolved population of young stars beyond the region of the dark clouds, which was extensively surveyed in the past.}

\maketitle

\section{Introduction}\label{section1}

In the early 1960s, \citet{Herbig1960} estimated the age of the two variable stars R~CrA and T~CrA associated with nebulosity based on the expected time required for them to contract to the main sequence ($\sim10^{7}$~yr) and showed that they were young. This encouraged astronomers to search for other young stars around these variables in the constellation of Corona-Australis. Indeed, subsequent studies revealed a wealth of young stellar objects (YSO) in this region from the most embedded protostars to the more evolved disc-free stars,  and Corona-Australis became one of the main targets for many studies related to star formation. 

The first optical and infrared surveys identified most of the hitherto known classical T~Tauri stars in Corona-Australis based on their strong H$\alpha$ and infrared excess emission \citep[see e.g.][]{Knacke1973,Glass1975,Marraco1981,Wilking1992,Wilking1997}. Later studies based on X-ray observations from the \textit{Einstein} Observatory \citep{Walter1986,Walter1997} and the ROSAT All-Sky Survey \citep{Neuhauser2000} identified many weak-line T~Tauri stars and a dispersed population of them surrounding the dark clouds of the region (the so-called off-cloud stars). The youngest YSOs in this region are the Class~0/I stars located in the Coronet cluster \citep{Taylor1984}. These sources have been monitored over the last decade based on multi-wavelength observations in order to characterise the properties of YSOs at this early stage of stellar evolution and confirm membership in the region \citep{Forbrich2006,Forbrich2007,Forbrich2007b}. So far, only a few brown dwarfs (and candidates) have been discovered in Corona-Australis and they are typically late M dwarfs \citep{Wilking1997,Fernandez2001,Bouy2004,Lopez-Marti2005}. 

In the most recent review, \citet{Neuhauser2008} compiled a list of 63 known YSOs identified in the literature that are likely to be associated to the Corona-Australis star-forming region. However, more recently \citet{Peterson2011} used infrared observations collected with the \textit{Spitzer Space Telescope} and identified new YSOs. The resulting list with 116~YSOs of that study almost doubled the number of known members in Corona-Australis and represents a major improvement to derive a complete census of the YSOs in this region.  

Although we have progressed in recent years to provide a more complete picture of the stellar content in Corona-Australis, the distance to it is still poorly constrained. Distances to individual stars are particularly important for YSOs to accurately derive their ages, masses, space motions, and confirm membership. \citet{Gaposchkin1936} and \citet{Marraco1981} estimated the distance towards Corona-Australis to be $150\pm50$~pc and 129~pc, respectively. The \textit{Hipparcos} satellite \citep{ESA1997} measured the trigonometric parallax of only five stars in Corona-Australis, but the resulting distances were mostly very imprecise and of minimal use. In the following year after publication of the \textit{Hipparcos} results,  \citet{Casey1998} inferred the distance of $129\pm11$~pc to the eclipsing binary system TY~CrA based on its orbital motion. Since then, most studies in the literature have adopted the distance of 130~pc to the Corona-Australis region. More recently, the first data release of the \textit{Gaia} space mission \citep[Gaia-DR1,][]{GaiaDR1} delivered trigonometric parallaxes of the following four stars in this region: RXJ1841.8-3525, RXJ1842.9-3532, CrAPMS~4SE, and HD~176386. The mean parallax of these stars ($\varpi=6.8\pm0.3$~mas) yields a distance of $146\pm6$~pc and suggests that the adopted distance to Corona-Australis needs to be revised. 

In this context, the second data release of the \textit{Gaia} space mission \citep[Gaia-DR2,][]{GaiaDR2} allowed us to search for additional members in Corona-Australis and revisit the distance to this region. Despite the highly variable extinction \citep[see e.g.][]{Cambresy1999,Dobashi2005,Alves2014}, which, in general, affects optical observations in Corona-Australis, one can still use the \textit{Gaia} data to search for additional members in the outskirts of the densest cloud cores as we explain here.   

This paper is one in a series dedicated to investigate open clusters and star-forming regions as part of the Dynamical Analysis of Nearby Clusters project \citep[DANCe,][]{Bouy2013}. In particular, the study of Corona-Australis will be divided into two parts. In this first paper, we report on the discovery of a distributed population of YSOs using only Gaia-DR2 data in an extended region around the molecular cloud complex. In a companion paper, we will use auxiliary data from the DANCe project to complement the Gaia-DR2 catalogue in a small region centred around the densest clouds, and we will use alternative methods to overcome the problem of extinction and search for additional members. The two studies combined together will deliver a complete census and the initial mass function of the Corona-Australis association. This paper is structured as follows. In Section~\ref{section2} we describe our membership analysis to search for new members in Corona-Australis based on the methodology previously developed by our team \citep{Sarro2014,Olivares2019}. Section~\ref{section3} is dedicated to the characterisation of the newly identified members in this study. We discuss the existence of substructures in the Corona-Australis region, compute distances and 2D velocities for individual stars from Bayesian inference, and classify the newly discovered members as Class~I, II, or III stars based on their infrared excess emission. Finally, we summarise our results and conclusions in Section~\ref{section4}.

\section{Membership analysis}\label{section2}

We present in this section our strategy to search for new members of the Corona-Australis star-forming region based on the algorithm developed by \citet{Sarro2014}, which was later modified by \citet{Olivares2019}. Briefly, the methodology models the field and cluster populations using Gaussian mixture models (GMM) in a representation space that takes proper motions, parallaxes, and multi-band photometry together with the corresponding uncertainties and correlations (when available). The field model is computed only once and fixed during the whole process, while the cluster model is built iteratively based on an initial list of cluster members given in the first iteration. The method assigns membership probabilities to the sources and classifies them into field stars and cluster members based on a probability threshold $p_{in, }$ which is predefined by the user. The resulting list of cluster members is used as input for the next iteration and the process is repeated until convergence. The solution is said to converge when the list of cluster members remains fixed after successive iterations. In the following, we describe the main steps of our membership analysis and we refer the reader to the original papers for more details about the methodology. 

\subsection{Initial list of stars in Corona-Australis} \label{section2.1}

The methodology that we use here starts with an initial list of cluster members in the first iteration to construct the cluster model that will be refined in the following iterations. This first list can be incomplete and somewhat contaminated since its main purpose is only to define the cluster locus in the space of parameters. We proceed as follows to construct the initial list of candidate stars in the Corona-Australis region. 

First, we compiled a list of known YSOs in this region that are published in the literature. We combined the sample of 63~stars given in Tables~1 and 2 of \citet{Neuhauser2008} with the list of 122 stars given in Tables~4, 5, 6, and 7 of \citet{Peterson2011}. Then, we cross-matched this list of stars with the Gaia-DR2 catalogue to retrieve the best astrometry available to date for our targets. This procedure uses the \texttt{TMASS\_BEST\_NEIGHBOUR} auxiliary table that is given in the Gaia archive and provides the Gaia-DR2 and 2MASS identifiers \citep{Cutri2003} of the sources that are in common between the two surveys. We used the 2MASS identifiers of our targets, which were known a priori the search for the corresponding Gaia-DR2 counterparts in this table and in order to avoid erroneous cross-matches. Then, we used the resulting Gaia-DR2 identifier of each source to retrieve its astrometry from the main catalogue table (\texttt{GAIA\_SOURCE}). We repeated this procedure for all sources with a 2MASS counterpart in our sample and searched the remaining ones in the Gaia-DR2 catalogue using their positions with a search radius of 1\arcsec. We find a one-to-one relationship for most sources in the sample, but we note that 2MASS~J19014055-3644320 and 2MASS~J19031185-3709020 have been resolved by the Gaia satellite. In such cases, we have kept the two components of the system in our sample. The list of stars compiled by \citet{Neuhauser2008} only includes the first binary system, which adds the number of entries of their list to 64~stars. The two binary systems are included in the samples of \citet{Peterson2011}, making it a total of 124 stars for that study.  After removing the 39 sources that are in common between the two studies, we ended up with a sample of 149 stars, which represents only a compilation of members (and candidate members) to the Corona-Australis region known in the literature  at this stage. We found proper motions and parallaxes in Gaia-DR2 for 87 stars of this initial sample following the strategy described above. 

Second, we refined the list of known YSOs and removed potential outliers based on Gaia-DR2 proper motions and parallaxes as well as objects with unreliable Gaia DR2 measurements. In this context, we used the re-normalised unit weight error (RUWE) criterion to remove the Gaia-DR2 sources in our sample with poor astrometric solutions (i.e. RUWE $\geq1.4$)\footnote{see technical note \href{https://www.cosmos.esa.int/web/gaia/dr2-known-issues}{GAIA-C3-TN-LU-LL-124-01} for more details}. After applying this selection criteria, our initial sample was reduced to 68 stars. To identify potential outliers in this sample, we computed robust distances, which are given by
\begin{equation}
RD(\mathbf{x})=\sqrt{(\mathbf{x}-\boldsymbol{\mu})^{t}\boldsymbol{\Sigma}^{-1}(\mathbf{x}-\boldsymbol{\mu})}\, ,
\end{equation} 
where $\boldsymbol{\mu}$ and $\boldsymbol{\Sigma}$ denote the multivariate location and covariance matrix obtained from the minimum covariance determinant \citep[MCD,][]{Rousseeuw1999} estimator. We used a $97.5\%$ tolerance ellipse to identify 16 sources in our sample as outliers based on their robust distances. The cutoff threshold to distinguish between cluster candidate members and potential outliers in our sample is given by $\sqrt{\chi^{2}_{p,\alpha}}$, where  $\chi^{2}_{p,\alpha}$ is the $\alpha$-quantile of the $\chi^{2}_{p}$ distribution. This preliminary analysis is based only on the 3D space of proper motions and parallaxes (i.e. $p=3$) and we used $\alpha=0.975$ to construct the tolerance ellipse. By doing so, we retain 52 known YSOs in our list as probable cluster members. 

Third, we searched for additional cluster candidate members in the Gaia-DR2 catalogue with proper motions and parallaxes that are similar to the known members in this region aiming to better constrain the cluster locus in the space of parameters with a more significant number of stars. We selected the Gaia-DR2 sources (after applying the RUWE criterion) that lie within the observed range of proper motion and parallax for membership in Corona-Australis (as defined from the sample of 52 YSOs). By doing so, we find 149 new cluster candidate members. By combining this list of stars with the 52 YSOs from the literature, we arrive at a sample of 201 stars that we use in the first iteration of our membership analysis.  


\subsection{Representation space}\label{section2.2}

The representation space is the set of observables that we used in the membership analysis to classify the sources as cluster members or field stars. It includes both the astrometric and photometric parameters given in the Gaia-DR2 catalogue. In general, proper motions and parallaxes are the most discriminant features to distinguish between the two populations. 
The three photometric bands ($G$, $G_{BP}$, $G_{RP}$)  given in Gaia-DR2 allowed us to construct colour-magnitude diagrams (CMD) using different combinations of them. We ran a random-forest classifier \citep[as described by][]{Olivares2019} to measure the relative importance of the photometric features (i.e. magnitudes and colours). This analysis suggests that $G_{RP}$ is the most important photometric feature; furthermore, $G_{BP}-G_{RP}$ and $G-G_{RP}$ are the most important colours to be included in our analysis. However, it should be noted that some inconsistencies in the blue (BP) photometric system have recently been reported in the literature \citep[see e.g.][]{Apellaniz2018}. Indeed, a preliminary membership analysis using CMDs based on the BP photometry showed a large spread for faint sources ($G\gtrsim 18$~mag), making our models less reliable when distinguishing between cluster members and field stars in this magnitude range. We have therefore decided to only work with the $G$ and $G_{RP}$ photometric bands. Thus, the representation space that we use here is defined by the observables $\mu_{\alpha}\cos\delta$, $\mu_{\delta}$, $\varpi$, $G_{RP,}$ and $G-G_{RP}$. 

\subsection{Field and cluster model}\label{section2.3}

To perform the membership analysis described in this paper, we downloaded the Gaia-DR2 catalogue in the region defined by
$0^{\circ}\leq l \leq 4^{\circ}$ and $-26^{\circ}\leq b \leq -10^{\circ}$\, as well as $356^{\circ}\leq l \leq 360^{\circ}$ and $-26^{\circ}\leq b \leq -10^{\circ}$, which clearly extends beyond the location of the Coronet cluster and known YSOs in the Corona-Australis region. In this region, we have a total of 12\,257\,645 sources in Gaia-DR2, after applying the RUWE selection criterion, and 10\,618\,999 sources with complete data in the chosen representation space. We constructed different models for the field population using GMM with 60, 80, 100, 120, 140, 160, and 180 components based on a random sample of $10^{6}$ sources, and we computed the Bayesian information criteria (BIC) for each one of them. We chose the GMM model with 100 components as the optimum model for the field population since it returns the smallest BIC value.  

The cluster model is the result of the two independent models for the astrometric and photometric features. The astrometric model is based on a GMM where the model parameters were inferred from the list of cluster members and the number of components was obtained from the BIC at each iteration. The photometric model used a multivariate Gaussian function of the photometric features in the chosen representation space to model the principal curve of the cluster (i.e. isochrone). Then, we computed the cluster and field likelihoods for each source and assigned Bayesian membership probabilities using, as prior, the fraction of sources in each category (member and non-members), which were obtained in the previous iteration. The sources are classified as members and non-members based on an internal probability threshold $p_{in}$ that is predefined by the user. This procedure was only applied to the sources with complete data in our representation space, which was used to train the model and update the list of members at each iteration. Once our solution converged, we generated a synthetic dataset and defined the optimum probability threshold $p_{opt}$ \citep[as described in Sect.~4.2.7 of][]{Olivares2019} to perform a final classification of all sources in the field into cluster members (i.e. prob. $\geq p_{opt}$) and non-members. The latter step includes sources with complete and incomplete data. 

\subsection{Final list of cluster members}\label{section2.4}

We ran the membership analysis as described in the previous sections by using different probability threshold values for $p_{in}$ (0.5, 0.6, 0.7, 0.8, and 0.9), and we compare our results in Table~\ref{tab_comp_pin}. As described in \citet{Olivares2019}, the contamination and recovery rates are estimated by performing the analysis with a synthetic sample of stars that mimic the cluster members. We defined two indicators to evaluate the quality of our solutions: the true positive rate (TPR, i.e. the fraction of cluster members generated in the synthetic datasets that are recovered by the algorithm) and the contamination rate (CR, i.e. the fraction of field stars generated in the synthetic datasets that are identified as cluster members by the algorithm). The high TPRs and low CRs given in Table~\ref{tab_comp_pin} for all the solutions confirm the robustness and consistency of our results that were obtained with different probability thresholds. However, we caution the reader in the sense that these values for the TPR and CR were obtained for synthetic datasets sampled from the inferred model and they cannot be understood as absolute measures for the true properties of the solution, but rather as estimates that can be computed in the absence of the true distributions. 

We note that 310 stars are in common among all the solutions in Table~\ref{tab_comp_pin}, which we obtained with different values for $p_{in}$. This shows that a very high fraction (i.e. 99$\%$) of the cluster members obtained with $p_{in}=0.8$ and $p_{in}=0.9$ were also recovered in other solutions, confirming them to be likely members of the Corona-Australis region. The results obtained with $p_{in}=0.9$ return a slightly lower CR (and higher TPR), so we conservatively adopt this solution (with 313~stars) as our final list of cluster members for the present study. Table~\ref{tab_members} lists the 313 members selected in our analysis and their properties derived in the following sections. In addition, we also provide the list of membership
probabilities for all the 10\,618\,999 sources in the field in Table~\ref{tab_prob} (using different values for $p_{in}$) so that the readers may select other cluster members with different constraints that are more specific to their scientific objectives. 

\begin{figure*}
\begin{center}
\includegraphics[width=0.33\textwidth]{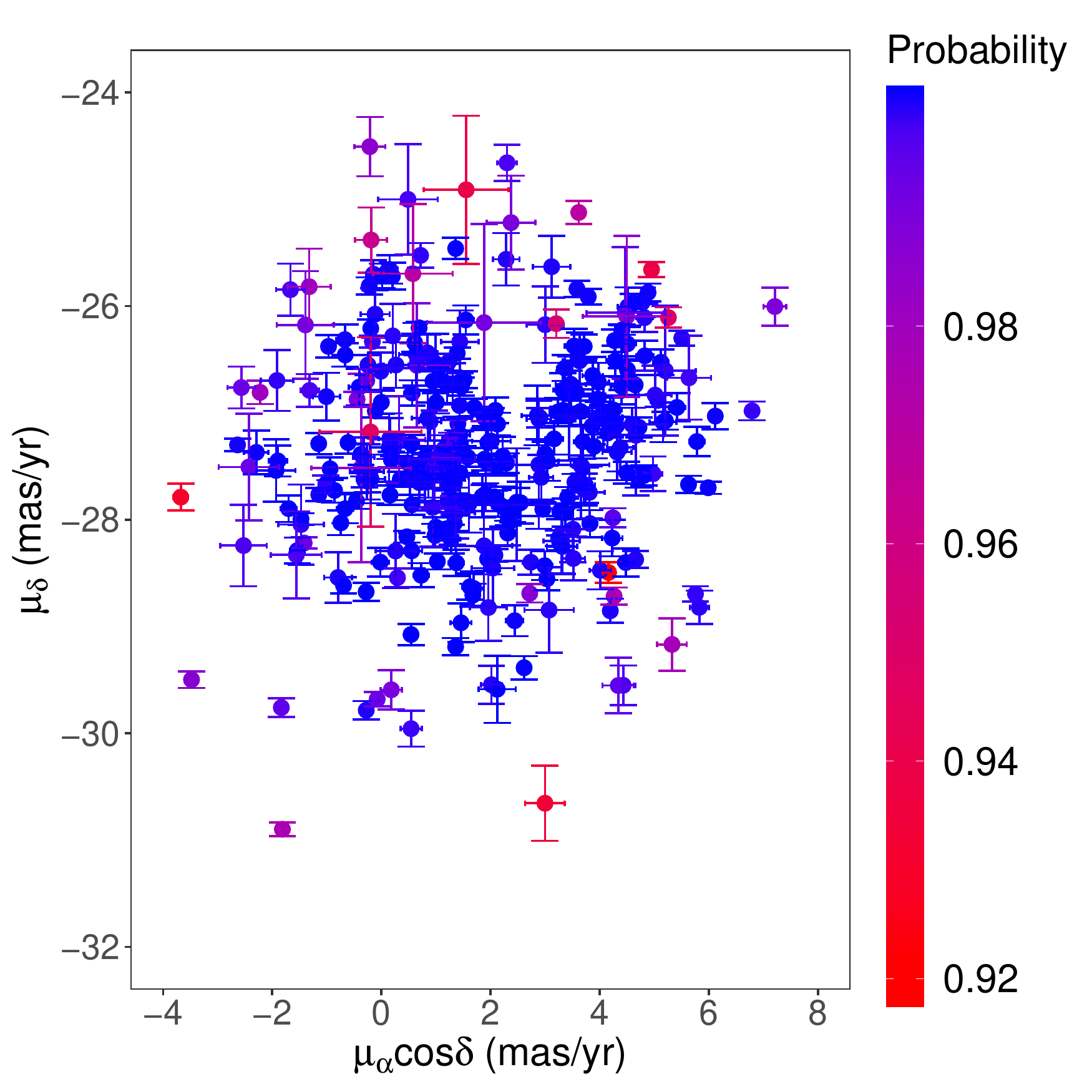}
\includegraphics[width=0.33\textwidth]{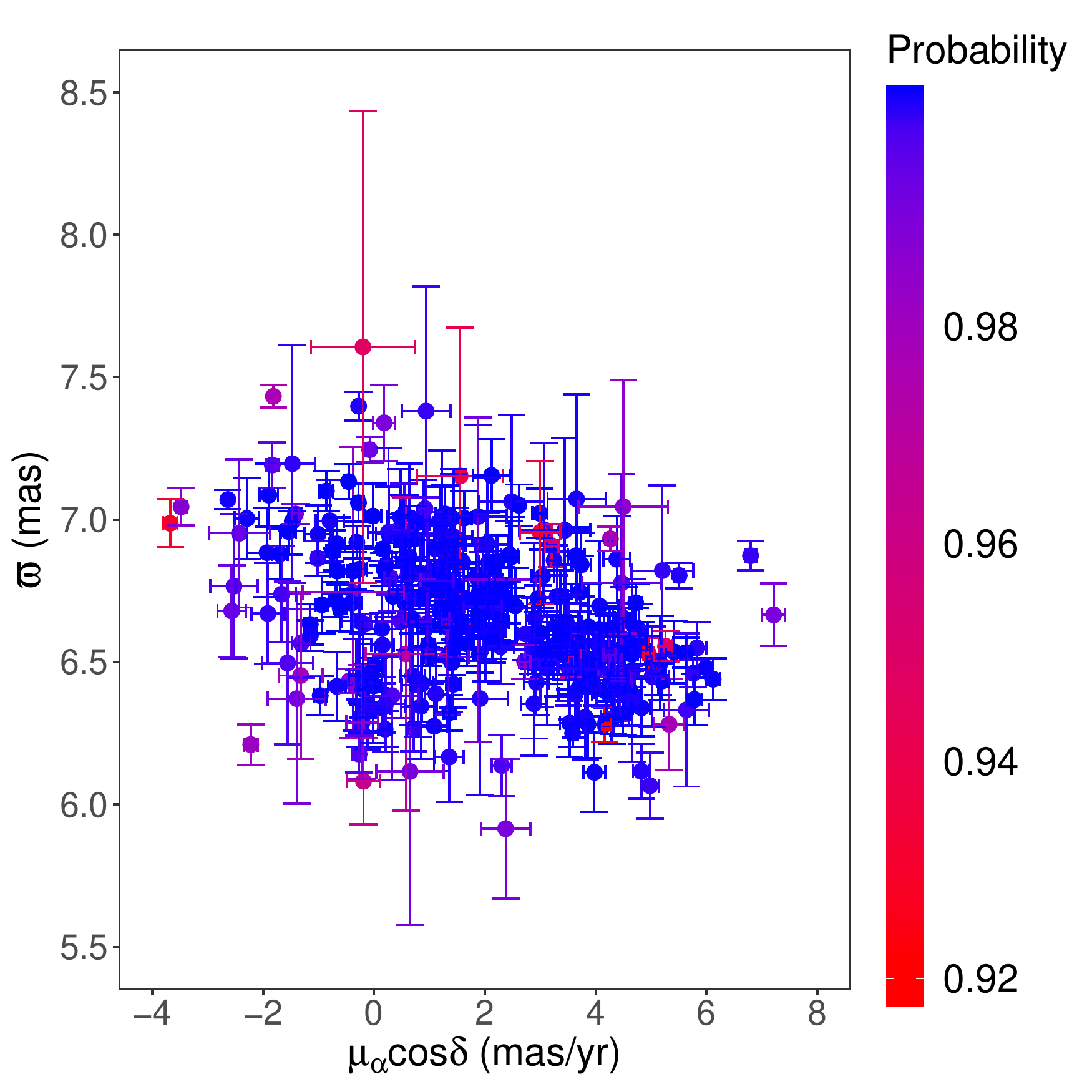}
\includegraphics[width=0.33\textwidth]{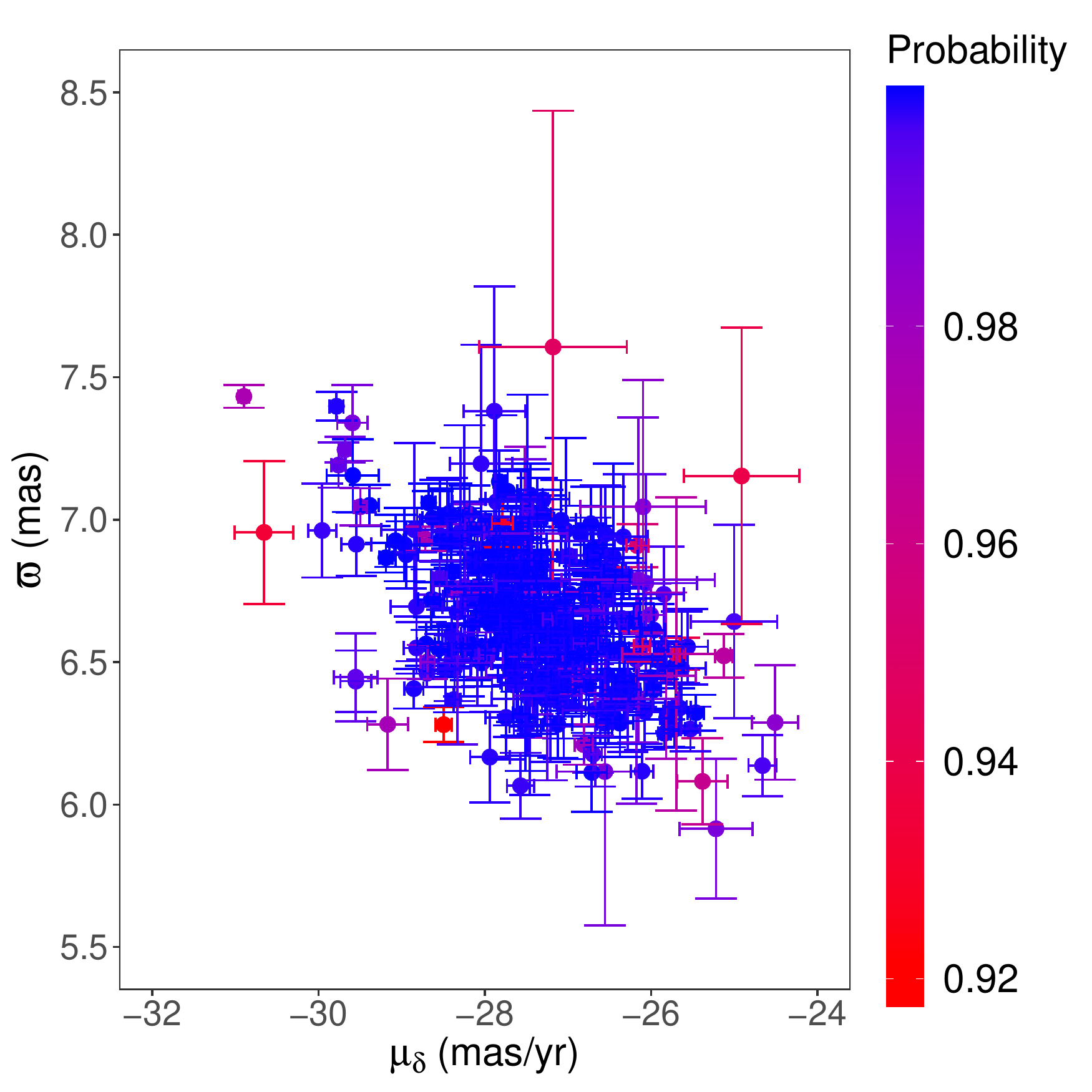}
\caption{
\label{fig_pmra_pmdec_parallax}
Proper motions and parallaxes of the 313 stars identified in our analysis as members of the Corona-Australis star-forming region. The stars are colour-coded based on their membership probabilities, which are scaled from zero to one. 
}
\end{center}
\end{figure*}

\begin{figure}
\begin{center}
\includegraphics[width=0.49\textwidth]{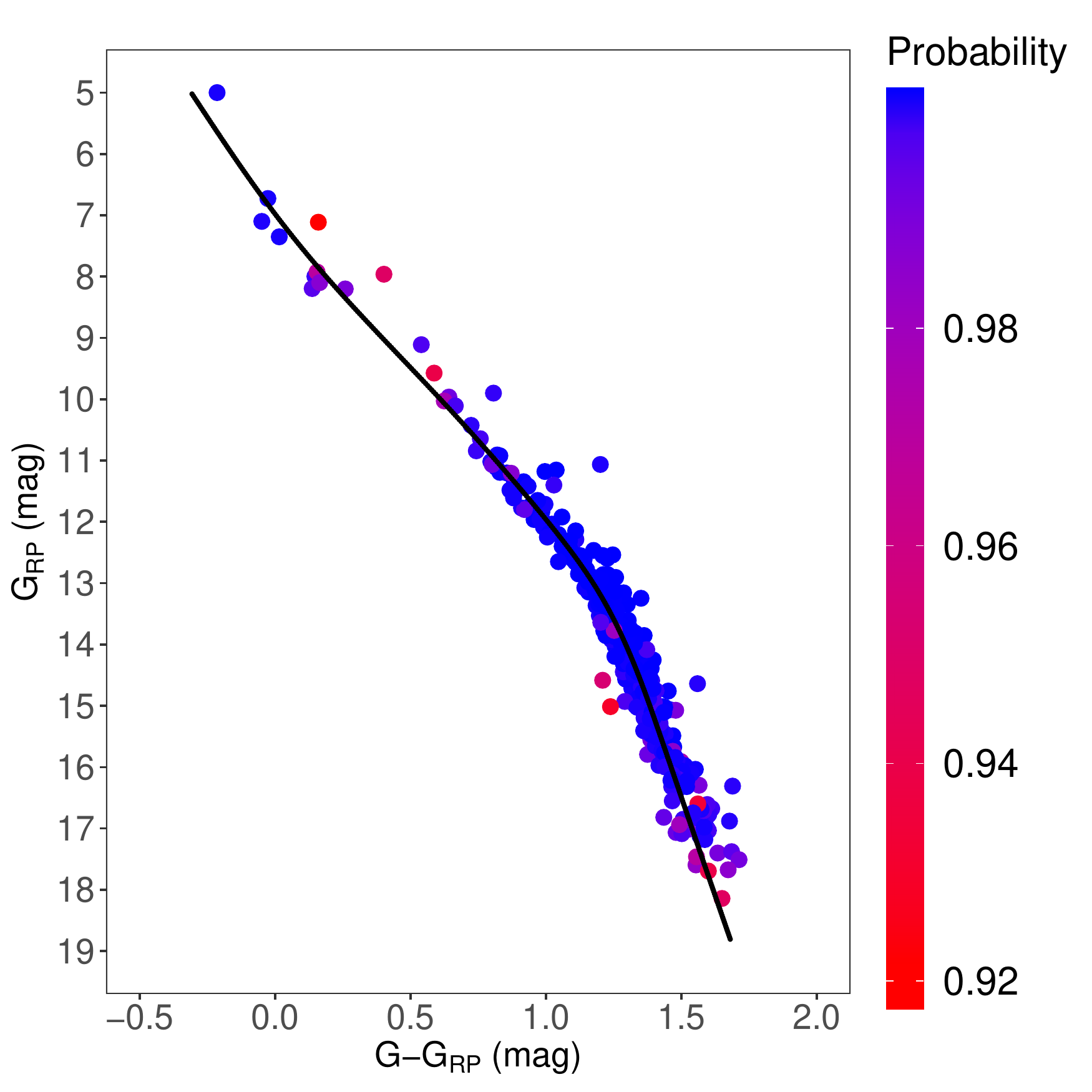}
\caption{
\label{fig_CMD}
Colour-magntiude diagram in the chosen representation space that was used to identify the 313 stars of the Corona-Australis star-forming region. The stars are colour-coded based on their membership probabilities, which are scaled from zero to one. The black line indicates the empirical isochrone that was derived from our analysis. 
}
\end{center}
\end{figure}

Figure~\ref{fig_pmra_pmdec_parallax} shows the cluster locus in the astrometric space of proper motions and parallaxes. As expected, stars with lower membership probabilities are mostly distributed in the outskirts of the proper motion and parallax distributions. The figure also shows the existence of substructures in our sample, which we discuss in more detail in Sect.~\ref{section3}. Figure~\ref{fig_CMD} shows the CMD in the chosen representation space and reveals the scarcity of early-type stars in our list of members. On the other hand, we note that our methodology allowed us, for the first time, to identify cluster members up to $G_{RP}\simeq 18$~mag in this region. The empirical isochrone that we obtained from our analysis is given in Table~\ref{tab_isochrone}. 

We note that the two variables R~CrA and T~CrA, which are often associated to the Corona-Australis region, are not included in our final list of members. R~CrA has a Gaia-DR2 parallax of $\varpi=10.536\pm0.697$~mas,  which is clearly inconsistent with other cluster members, and its proper motion ($\mu_{\alpha}\cos\delta=1.582\pm1.196$~mas/yr and $\mu_{\delta}=-30.835\pm1.193$~mas/yr) would place it only in the outskirts of the observed distribution of proper motion defined by other cluster members (see Fig.~\ref{fig_pmra_pmdec_parallax}). Previous results from the new reduction of the Hipparcos catalogue \citep{HIP2007} delivered proper motion ($\mu_{\alpha}\cos\delta=-28.30\pm42.68$~mas/yr and $\mu_{\delta}=20.57\pm22.97$~mas/yr) and parallax ($\varpi=40.93\pm27.95$~mas) measurements, which were not precise enough to draw firm conclusions, but they already suggested that R~CrA was not a member of the Corona-Australis association based on its astrometry. The UCAC5 \citep{UCAC5} proper motion of R~CrA measured from the ground  ($\mu_{\alpha}\cos\delta=7.7\pm1.2$~mas/yr and $\mu_{\delta}=-17.6\pm1.2$~mas/yr) is also inconsistent with membership in Corona-Australis. In addition, its radial velocity of $V_{r}=-36.0\pm4.9$~km/s \citep{Gontcharov2006} significantly exceeds the observed radial velocity for other cluster members \citep[][see also discussion in Sect.~\ref{section3.2}]{James2006}. Altogether, this explains the reason why R~CrA was rejected in our membership analysis. On the other hand, the Gaia-DR2 catalogue provides nor proper motion or parallax for T~CrA. This star was not observed by the Hipparcos satellite and it is also not listed in the UCAC5 catalogue. The former UCAC4 catalogue \citep{UCAC4} provides a proper motion result ($\mu_{\alpha}\cos\delta=2.0\pm3.8$~mas/yr and $\mu_{\delta}=-22.6\pm3.8$~mas/yr), which is consistent with membership in Corona-Australis (within the large uncertainties of that solution), but a parallax measurement would still be required to unambiguously confirm its membership status. T~CrA is not included in our list of members because we only used the Gaia-DR2 sources with complete astrometry in our membership analysis. The brightest star in our sample is HD~172910 (Gaia~DR2 6733635914056263296), a B2-type star \citep[see e.g.][]{Cucchiaro1980}, which was not listed as a member of the Corona-Australis association before this study and which might be the most massive and brightest member of the association. 

\begin{figure}
\begin{center}
\includegraphics[width=0.47\textwidth]{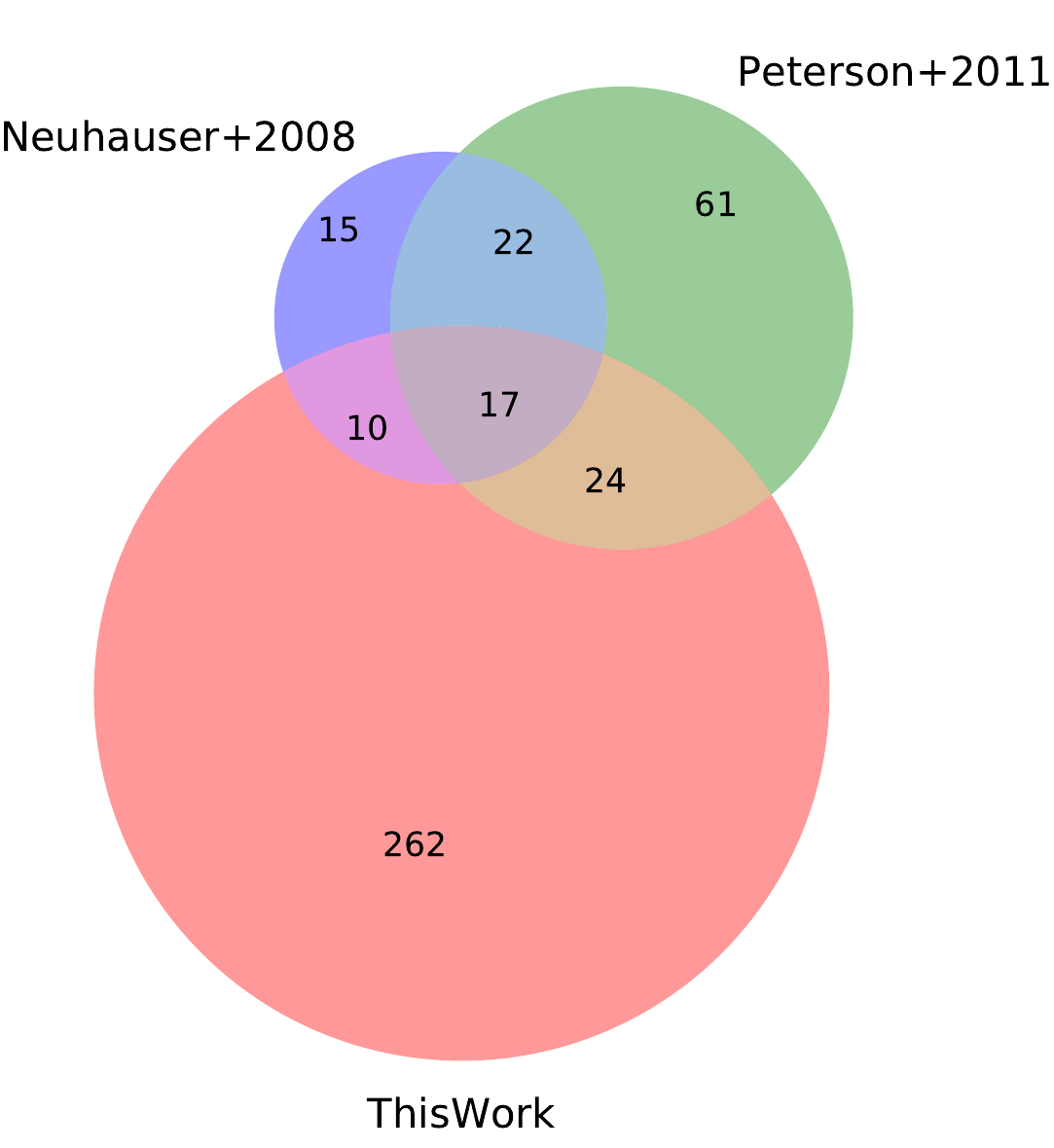}
\caption{
\label{fig_Venn}
Venn-diagram comparing the number of stars in common between our analysis and previous studies. The area of each shape is proportional to the number of stars in the corresponding sample.
}
\end{center}
\end{figure}

We verified that 180 stars from our initial list of 201 sources (see Sect.~\ref{section2.1}) have been confirmed as cluster members. The Venn-diagram shown in Figure~\ref{fig_Venn} illustrates the number of stars in our solution that are in common with previous studies in the literature. When counting the number of stars in each sample, it should be noticed that the samples from \citet{Neuhauser2008} and \citet{Peterson2011} add up to 64 and 124 stars, respectively (instead of 63 and 122 stars), because of the sources that have been resolved by the Gaia satellite as explained in Sect.~\ref{section2.1}. The membership analysis performed in this study allowed us to confirm 51 stars from the literature as cluster members. The remaining candidate members from the literature (with available astrometry), which were rejected by our analysis, have proper motions and/or parallaxes in Gaia-DR2 that are inconsistent with membership in Corona-Australis, and they lie below or above the empirical isochrone defined by the cluster members. In addition, we identify another 262 stars that are associated to the Corona-Australis star-forming region. This result increases the number of confirmed cluster members in this region by a factor of about 5. 

\subsection{Internal validation}\label{section2.5}

\begin{figure}
\begin{center}
\includegraphics[width=0.49\textwidth]{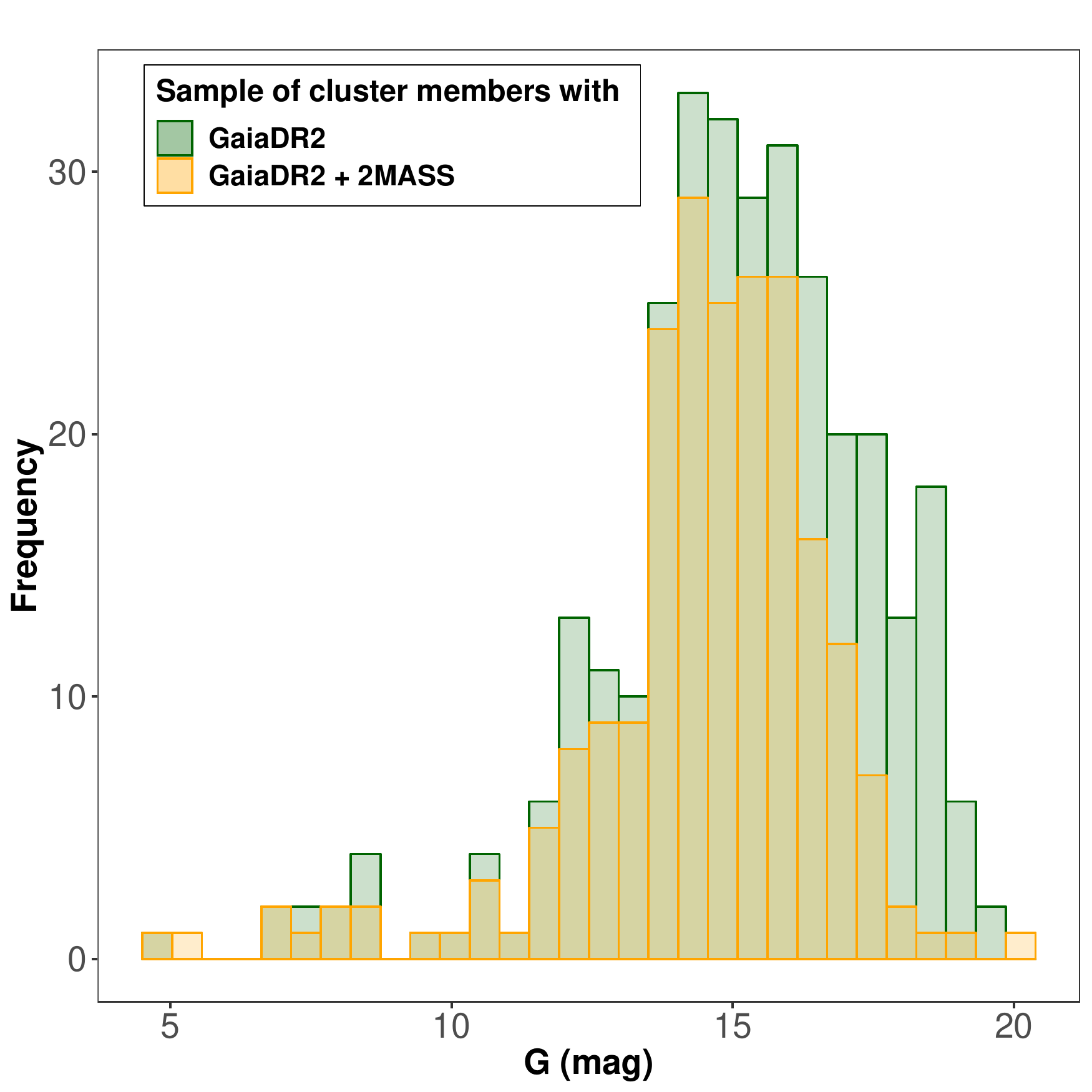}
\caption{
\label{fig_histGmag}
Histogram of magnitudes (G-band) of the cluster members selected based on two different representation spaces.  
}
\end{center}
\end{figure}

We repeated the membership analysis described in the previous sections using a different representation space to assess the robustness of our results. To increase the number of photometric features in our analysis we cross-matched the Gaia-DR2 and 2MASS catalogues in the region of the sky defined in Sect.~\ref{section2.3}. After running the random-forest classifier, we conclude that $K_{s}, H, G, J-H,$ and $G_{RP}-H$ are the most important photometric features to be included in our analysis. Thus, we ran a new membership analysis using the representation space defined by $\mu_{\alpha}\cos\delta$, $\mu_{\delta}$, $\varpi$, $K_{s}$, $H$, $G$, $J-H,$ and $G_{RP}-H$ with the same initial list of stars as before. By doing so, we found a sample of 216 cluster members using $p_{in}=0.9$. We note that 211 stars (i.e. 98$\%$ of the sample) are in common with the sample of 313 members obtained using only Gaia-DR2 data. This shows good agreement between the two solutions derived from different representation spaces. The smaller number of members identified in this alternative solution is explained by the shallower depth of the 2MASS catalogue. Figure~\ref{fig_histGmag} indeed shows that the faintest cluster members included in the Gaia-DR2 solution cannot be recovered by this model because our methodology only uses the sources with 2MASS photometry to construct the cluster model. We therefore prefer the solution given in Sect.~\ref{section2.4}, using only Gaia-DR2 data, which returns a more complete (deeper) census of the Corona-Australis region.

\begin{table}
\renewcommand\thetable{1} 
\centering
\caption{Comparison of our results for the membership analysis using different values for the probability threshold $p_{in}$. We provide the optimum probability threshold, the number of cluster members, the true positive rate (TPR), and contamination rate (CR) obtained for each solution. 
\label{tab_comp_pin}}
\begin{tabular}{ccccc}
\hline\hline
$p_{in}$&$p_{opt}$&Members&TPR&CR\\
\hline\hline
0.5&0.889 &326& $0.996 \pm 0.002$ & $0.009 \pm 0.002$\\
0.6&0.909 &317& $0.997 \pm 0.001$ & $0.006 \pm 0.002$\\ 
0.7&0.879 &322&$0.996 \pm 0.001$ & $0.006 \pm 0.003$\\ 
0.8&0.929 &312&$0.996 \pm 0.002$ & $0.007 \pm 0.003$\\ 
0.9&0.879 &313 &$0.997\pm 0.002$ & $0.006 \pm 0.001$\\ 
\hline\hline
\end{tabular}
\end{table}

\section{Discussion}\label{section3}

\subsection{Evidences of multiple stellar populations}

\begin{figure*}
\begin{center}
\includegraphics[width=0.33\textwidth]{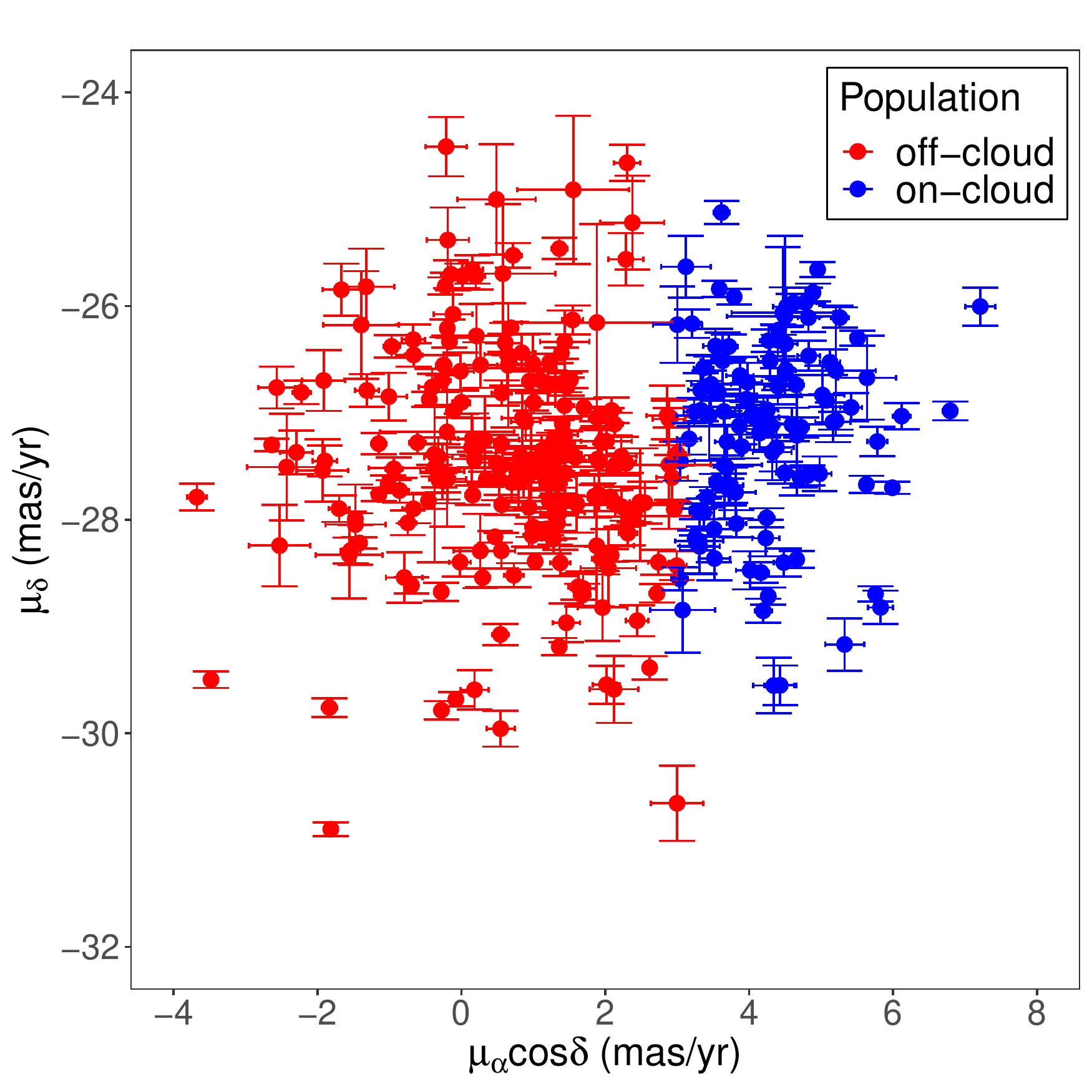}
\includegraphics[width=0.33\textwidth]{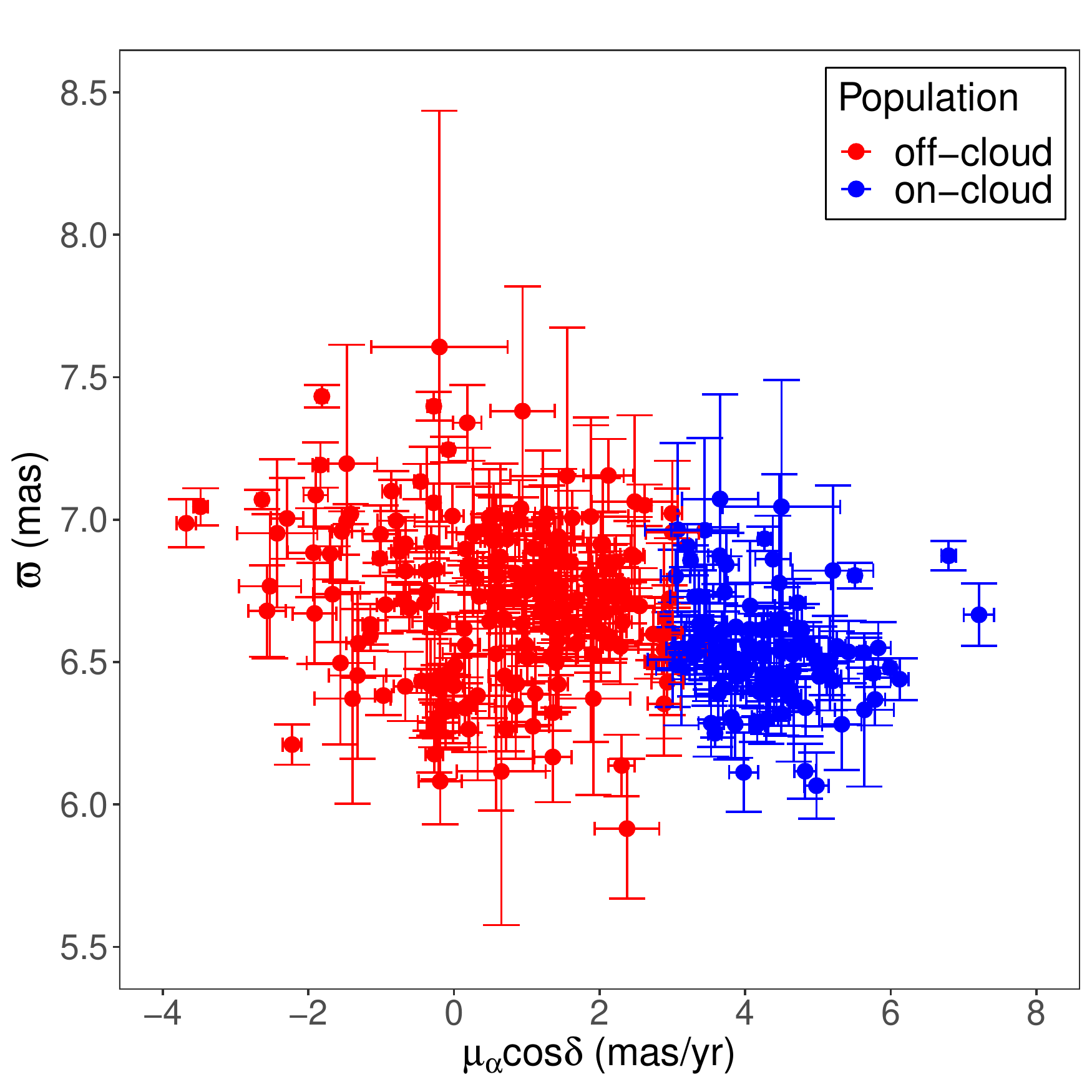}
\includegraphics[width=0.33\textwidth]{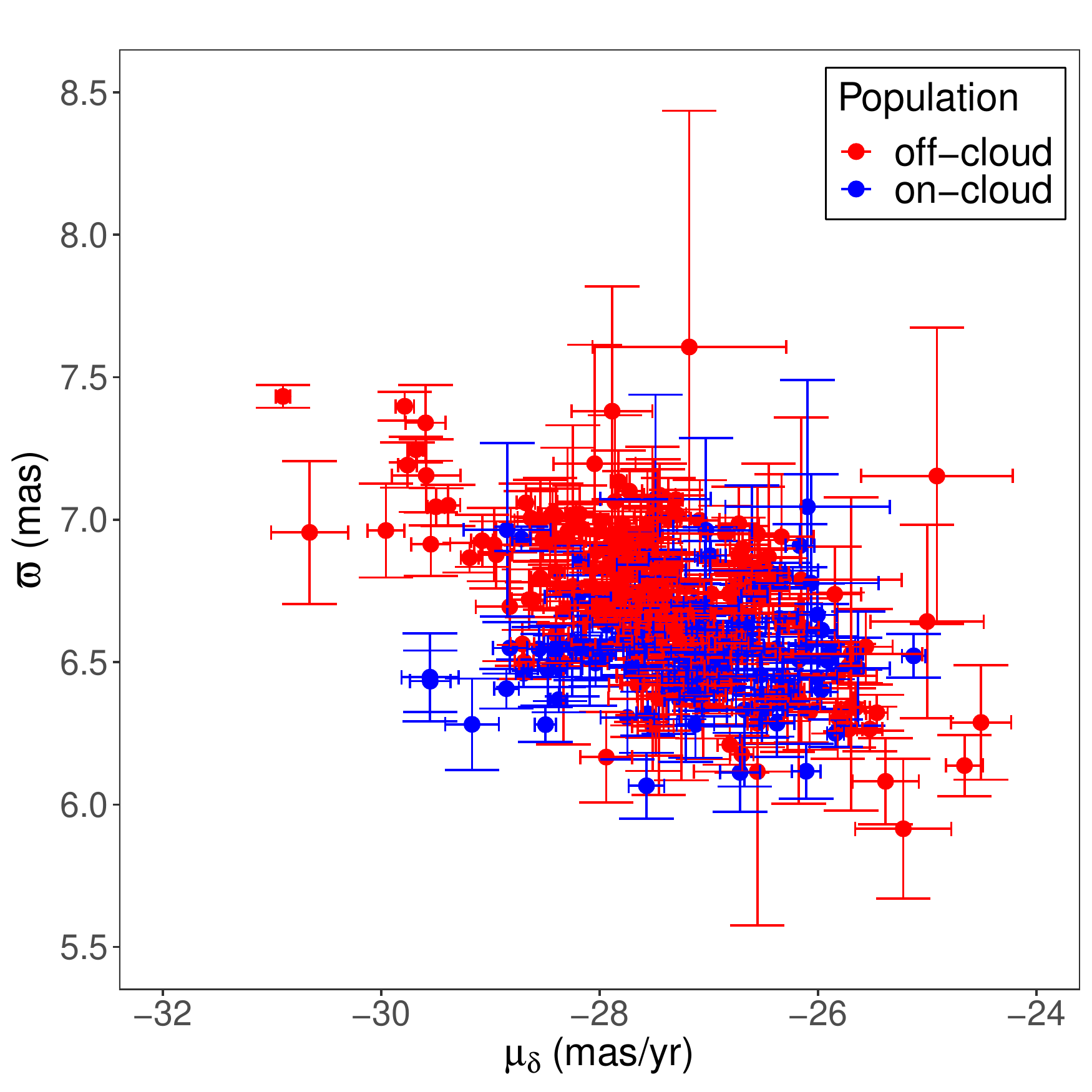}
\caption{
\label{fig_pmra_pmdec_parallax_groups}
Proper motions and parallaxes of the two subgroups of stars in our sample of cluster members. 
}
\end{center}
\end{figure*}

One interesting point that arises from our analysis is the existence of substructures (i.e. subgroups) in our sample of cluster members as already anticipated in Sect.~\ref{section2.4}. It is apparent from Figure~\ref{fig_pmra_pmdec_parallax} that the stars in our sample can be visually separated into two subgroups. The most discriminant feature in the astrometric space of observables is the proper motion component in right ascension and the borderline between the subgroups is located at about $\mu_{\alpha}\cos\delta\simeq3$~mas/yr. To better illustrate this discussion, in Figure~\ref{fig_pmra_pmdec_parallax_groups} we present the distribution of proper motions and parallaxes of the stars as done in Figure~\ref{fig_pmra_pmdec_parallax}, but we visually split the sample into these two subgroups. We assigned 106~stars with $\mu_{\alpha}\cos\delta>3$~mas/yr to one subgroup, and the remaining 207~stars to the other subgroup. Table~\ref{tab_subgroups} lists the mean proper motions and parallaxes of the two subgroups that we find in our sample. 

\begin{figure*}
\begin{center}
\includegraphics[width=1.0\textwidth]{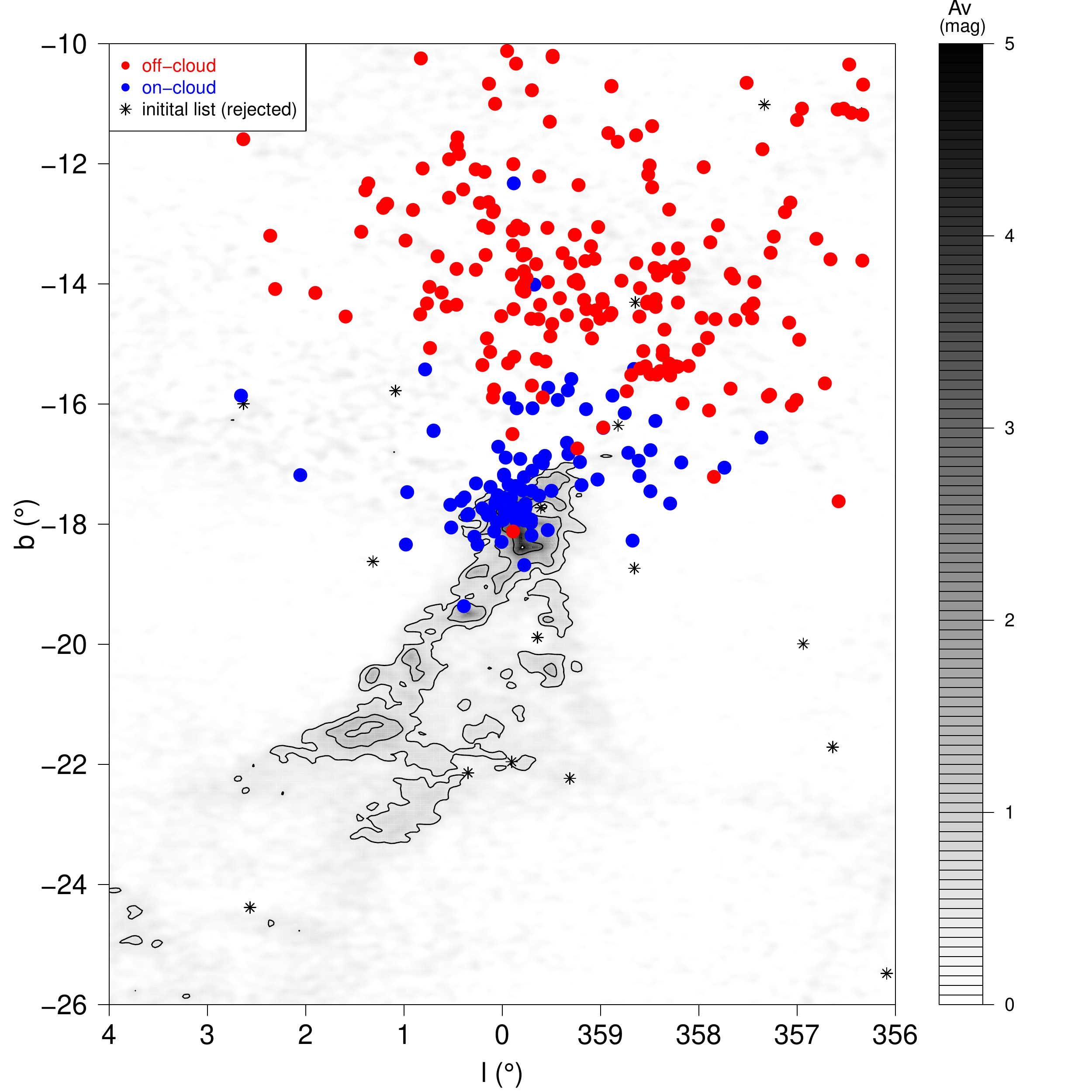}
\caption{
\label{fig_extmap}
Location of the 313 cluster members of the Corona-Australis star-forming region overlaid on the extinction map of \citet{Dobashi2005} in Galactic coordinates. Red and blue symbols indicate the off-cloud and on-cloud populations of stars, respectively. The black asterisks indicate the stars in our initial list of the membership analysis that have been rejected in this study. }
\end{center}
\end{figure*}

Figure~\ref{fig_extmap} shows that most of the stars in the first subgroup are located in a region of highly variable extinction that contains the dark clouds of the Corona-Australis region at its core (hereafter, the on-cloud population). This is the classical region that was surveyed by previous studies to search for new YSOs. The second subgroup of stars includes the more dispersed cluster members in our sample, which clearly extend beyond the main cores of gas and dust in this region (hereafter, the off-cloud population). \citet{Neuhauser2000} used X-ray observations from the ROSAT satellite and ground-based follow-up spectroscopy to detect a number of off-cloud weak-line T~Tauri stars in this region. The off-cloud population that we identify in our study based on Gaia-DR2 data greatly exceeds the sample of off-cloud stars reported in that paper, and it confirms the existence of such a dispersed population of young stars in the Corona-Australis star-forming region. It is interesting to note that the off-cloud population is restricted to the northern part of the Corona-Australis region and we did not detect any cluster member below $b\simeq-20^{\circ}$.

We performed a two sample Kolmogorov-Smirnov (KS) and Anderson-Darling (AD) test to quantitatively assess whether the two populations of cluster members in our sample exhibit the same (or different) proper motion and parallax distributions.  Our results are given in Table~\ref{tab_kstest}. By adopting a significance level of $\alpha=0.05,$ for example, we indeed conclude that the two populations exhibit different proper motion and parallax distributions. We therefore confirm the existence of multiple populations of stars associated to the Corona-Australis star-forming region, which were not known before this study. 

In a recent study, \citet{Gagne2018} discussed the existence of a stellar group, with ten members, in the vicinity of Corona-Australis, which the authors named Upper Corona-Australis (UCRA). We note that the following five stars of that sample are in common with the off-cloud population reported in this paper: HIP~92188, RX~J1839.0-3726, HD~172910, RX~J1842.9-3532, and  RX~J1841.8-3525. One star, namely RX~J1852.3-3700, was assigned to the on-cloud population of our study due to the group splitting in the space of proper motions as described above. The following three sources from that sample, RX J1844.3-3541, RX J1845.5-3750, and RX J1917.4-3756, were discarded from our analysis based on the RUWE selection criterion (see Sect.~\ref{section2.1}). Lastly, RX J1853.1-3609 was not included in our analysis as we found no Gaia-DR2 counterpart within 5\arcsec. Most of the UCRA members presented by \citet{Gagne2018} are indeed associated with the off-cloud population of Corona-Australis stars discussed in our study. We therefore argue that these UCRA group members belong to the much more numerous and extended population of YSOs in the north of the Corona-Australis dark clouds, which we discuss in this paper. 

\begin{table*}
\centering
\scriptsize{
\caption{Properties of the subgroups in Corona-Australis.
\label{tab_subgroups}}
\begin{tabular}{lcccccccccc}
\hline\hline
Sample&$N_{stars}$&\multicolumn{3}{c}{$\mu_{\alpha}\cos\delta$}&\multicolumn{3}{c}{$\mu_{\delta}$}&\multicolumn{3}{c}{$\varpi$}\\
&&\multicolumn{3}{c}{(mas/yr)}&\multicolumn{3}{c}{(mas/yr)}&\multicolumn{3}{c}{(mas)}\\
\hline\hline
&&Mean$\pm$SEM&Median&SD&Mean$\pm$SEM&Median&SD&Mean$\pm$SEM&Median&SD\\
\hline
Off-cloud & 207 & $ 0.654 \pm 0.096 $& 0.848 & 1.375 & $ -27.440 \pm 0.072 $& -27.472 & 1.038 & $ 6.757 \pm 0.019 $& 6.761 & 0.270 \\
On-cloud & 106 & $ 4.277 \pm 0.082 $& 4.248 & 0.843 & $ -27.183 \pm 0.088 $& -27.042 & 0.907 & $ 6.574 \pm 0.018 $& 6.561 & 0.190 \\
\hline
Full sample & 313 & $ 1.881 \pm 0.119 $& 1.604 & 2.107 & $ -27.353 \pm 0.057 $& -27.383 & 1.001 & $ 6.695 \pm 0.015 $& 6.673 & 0.260 \\
\hline\hline
\end{tabular}
\tablefoot{We provide for each subgroup the number of stars, mean, standard error of the mean (SEM), median and standard deviation (SD) of proper motions and parallaxes.}
}
\end{table*}

\begin{table}
\centering
\caption{Results of the KS and AD statistical tests applied to the distribution of proper motions and parallaxes of the on-cloud and off-cloud populations.
\label{tab_kstest}}
\begin{tabular}{lcc}
\hline\hline
&KS-test ($p$-value)&AD-test ($p$-value)\\
\hline
$\mu_{\alpha}\cos\delta$&$2.20\times10^{-16}$&$8.24\times10^{-63}$\\
$\mu_{\delta}$&$5.73\times10^{-4}$&$6.81\times10^{-3}$\\
$\varpi$&$5.43\times10^{-13}$&$1.85\times10^{-12}$\\
\hline\hline
\end{tabular}
\end{table}

\subsection{Distance and kinematics of Corona-Australis stars}\label{section3.2}

The new sample of cluster members, which were identified in this study from Gaia-DR2 data, allowed us to put firm constraints on the distance to the Corona-Australis star-forming region. We proceeded as follows to convert the parallaxes of individual stars into distances. 

First, we corrected the Gaia-DR2 parallaxes by the zero-point shift of -0.030~mas, which is present in the published data, and added 0.1~mas and 0.1~mas/yr in quadrature to the parallax and proper motion uncertainties to take the systematic errors of the Gaia-DR2 catalogue into account \citep[see e.g.][]{Lindegren2018}. This procedure does not affect our membership analysis presented in Sect.~\ref{section2} since it was applied to all sources in the field, but it needs to be considered when estimating distances and velocities. Second, we used Bayesian inference to convert the parallaxes and proper motions of the stars into distances and 2D tangential velocities. In this context, we used the exponentially decreasing space density prior for the distance with a length scale of $L=1.35$~kpc \citep{Bailer-Jones2015,Astraatmadja2016} and the beta function for the prior over speed following the online tutorials available in the Gaia archive \citep[see e.g.][]{Luri2018}\footnote{see also  \href{https://github.com/agabrown/astrometry-inference-tutorials/blob/master/3d-distance/resources/3D_astrometry_inference.pdf}{GAIA-C8-TN-LU-MPIA-CBJ-081} for more details.}. The resulting distances and tangential velocities that we derived for individual stars are given in Table~\ref{tab_members}. The distances range from $141.6_{-6.6}^{+9.1}$~pc to $164.2_{-3.9}^{+4.5}$~pc for the on-cloud population. The stars in the off-cloud population are more dispersed not only in an angular extent but also along the line of sight: the closest and remotest stars are located at $134.1_{-1.9}^{+1.9}$~pc and $168.3_{-6.5}^{+8.7}$~pc, respectively. 

\begin{figure}
\begin{center}
\includegraphics[width=0.49\textwidth]{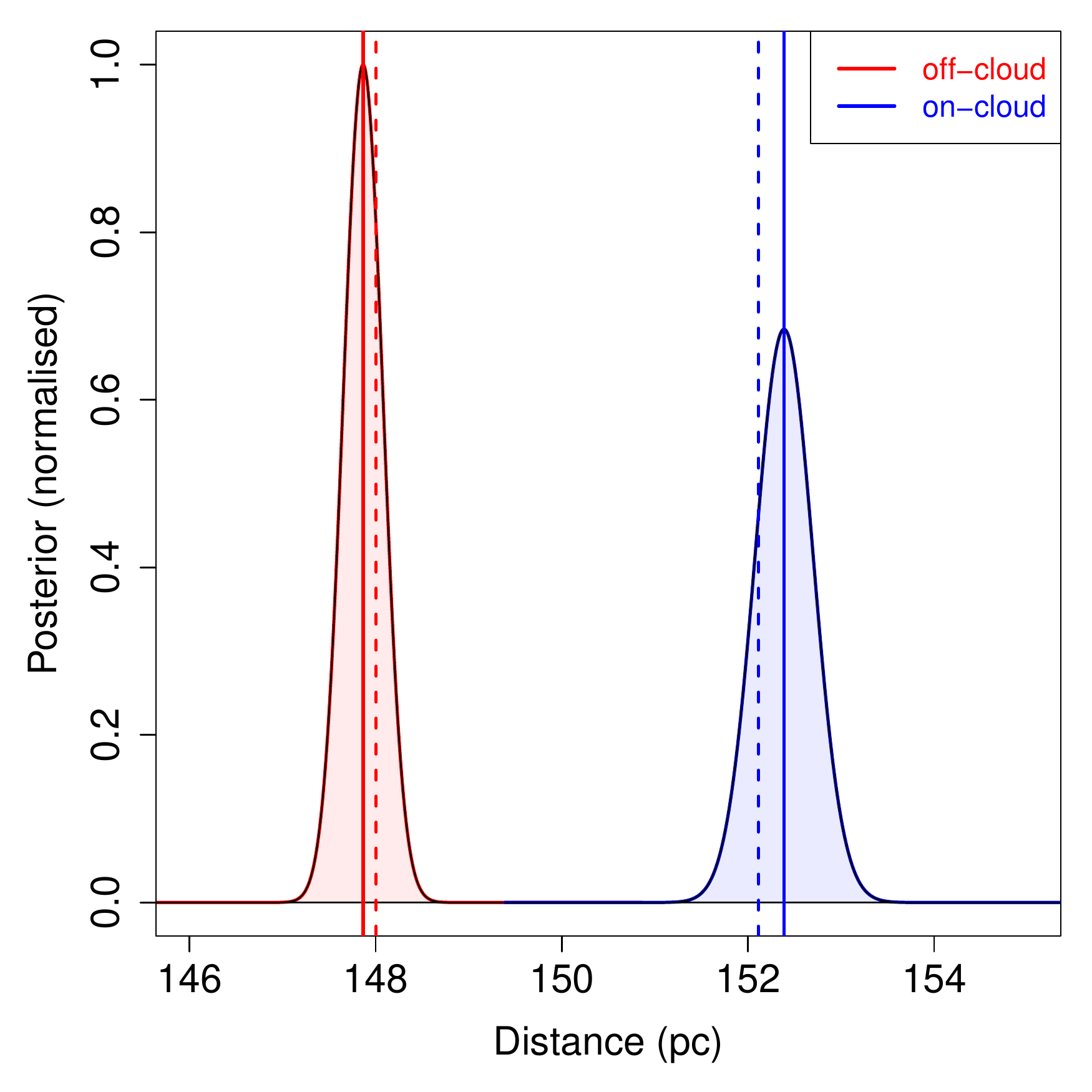}
\caption{
\label{fig_distance_subgroups}
Posterior probability density function of the distance to the two populations of cluster members in Corona-Australis. The solid and dashed lines indicate the distance estimates derived from the Bayesian approach and the inverse of the mean parallax, respectively. 
}
\end{center}
\end{figure}

Analogously, we computed the Bayesian distance estimate for each population of stars in our sample by using the online tutorials available in the Gaia archive to infer the distance to clusters \citep[see e.g.][]{Luri2018}. We proceeded in a similar manner as explained above for the case of a single star, but by using a multivariate likelihood that is the product of $N$ 1D Gaussians (where $N$ is the number of stars). This procedure took the same prior over distance as mentioned before and the resulting distances are given in Table~\ref{tab_distance}. The posterior probability function obtained from the Bayesian approach is illustrated in Figure~\ref{fig_distance_subgroups}. At this stage, we would like to mention that the exponentially decreasing space density prior used in this study for the distance has been proposed in the literature in the context of large samples with very wide distribution of parallaxes and uncertainties. Our sample is much more restricted in both parallax and uncertainty so that a more specific prior in our case would be recommended. However, thanks to the good precision of the Gaia-DR2 parallaxes in Corona-Australis (i.e. relative errors of about $1\%$), our results presented here do not differ significantly as compared to other priors. Our team is currently developing alternative priors for open cluster and young stellar associations (Olivares et al., in prep.), and we will soon be able to improve distance estimates to such stellar groups.

The distance estimate that we derive in this study for the off-cloud and on-cloud populations are $147.9^{+0.3}_{-0.4}$~pc and $152.4^{+0.4}_{-0.4}$~pc, respectively, which implies a distance variation of $4.5\pm0.1$~pc along the line of sight between the subgroups. Even though the two populations exhibit slightly distinct properties (e.g. in the proper motion component in right ascension), they are very close to each other and are part of the same star-forming complex. The distance estimate that we derived, which took all the 313 stars at once in the solution, is $149.4^{+0.4}_{-0.4}$~pc (see Table~\ref{tab_distance}).

Recent studies in the literature have reported other values for the zero-point correction of the Gaia-DR2 parallaxes by using different samples of stars and methods to derive this offset \citep[see][]{Kounkel2018,Riess2018,Stassun2018,Graczyk2019,Schoenrich2019,Zinn2019}. These values range from $-0.031\pm0.011$~mas \citep{Graczyk2019} to $-0.082\pm0.033$~mas \citep{Stassun2018}. The lower limit confirms the nominal zero-point shift derived by the Gaia team \citep{Lindegren2018}, which is used throughout our analysis. By applying the largest zero-point correction reported in the literature, we find a distance of $148.3^{+0.4}_{-0.3}$~pc with all the 313 stars of our sample. On the other hand, if we were not to have corrected the Gaia-DR2 parallaxes for any of the previously listed values (i.e. no zero-point correction), the distance that we would have derived for the full sample of stars would have become $150.1^{+0.4}_{-0.4}$~pc. Both results are consistent with the solution of $149.4^{+0.4}_{-0.4}$~pc, which was previously derived within the corresponding error bars. Therefore, we conclude that the distance inferred in this study based on Gaia-DR2 parallaxes exceeds, by about 20~pc, the canonical distance of 130~pc that is commonly used in the literature for the Corona-Australis region. This conclusion is independent of the zero-point correction that we use.

\begin{figure*}
\begin{center}
\includegraphics[width=0.49\textwidth]{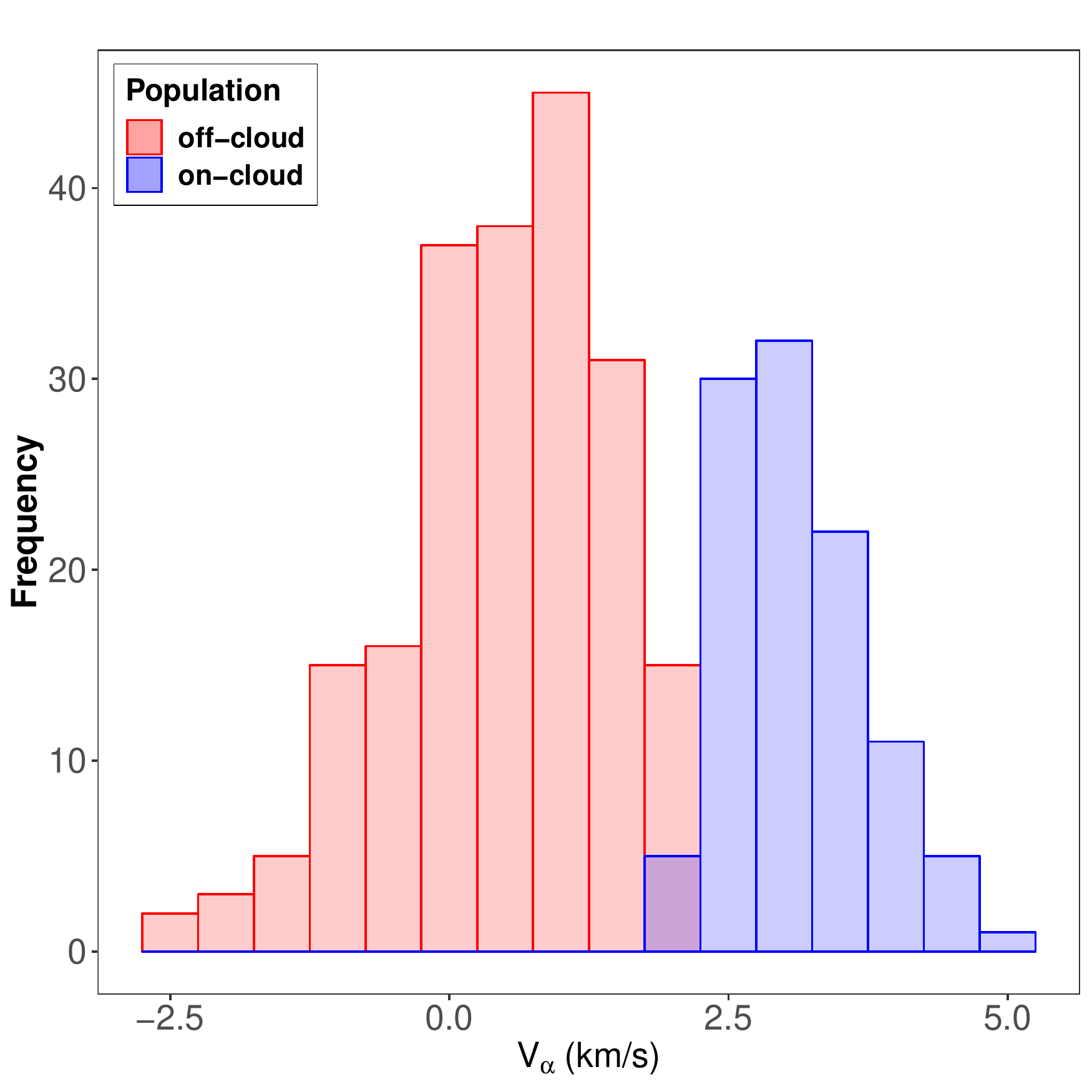}
\includegraphics[width=0.49\textwidth]{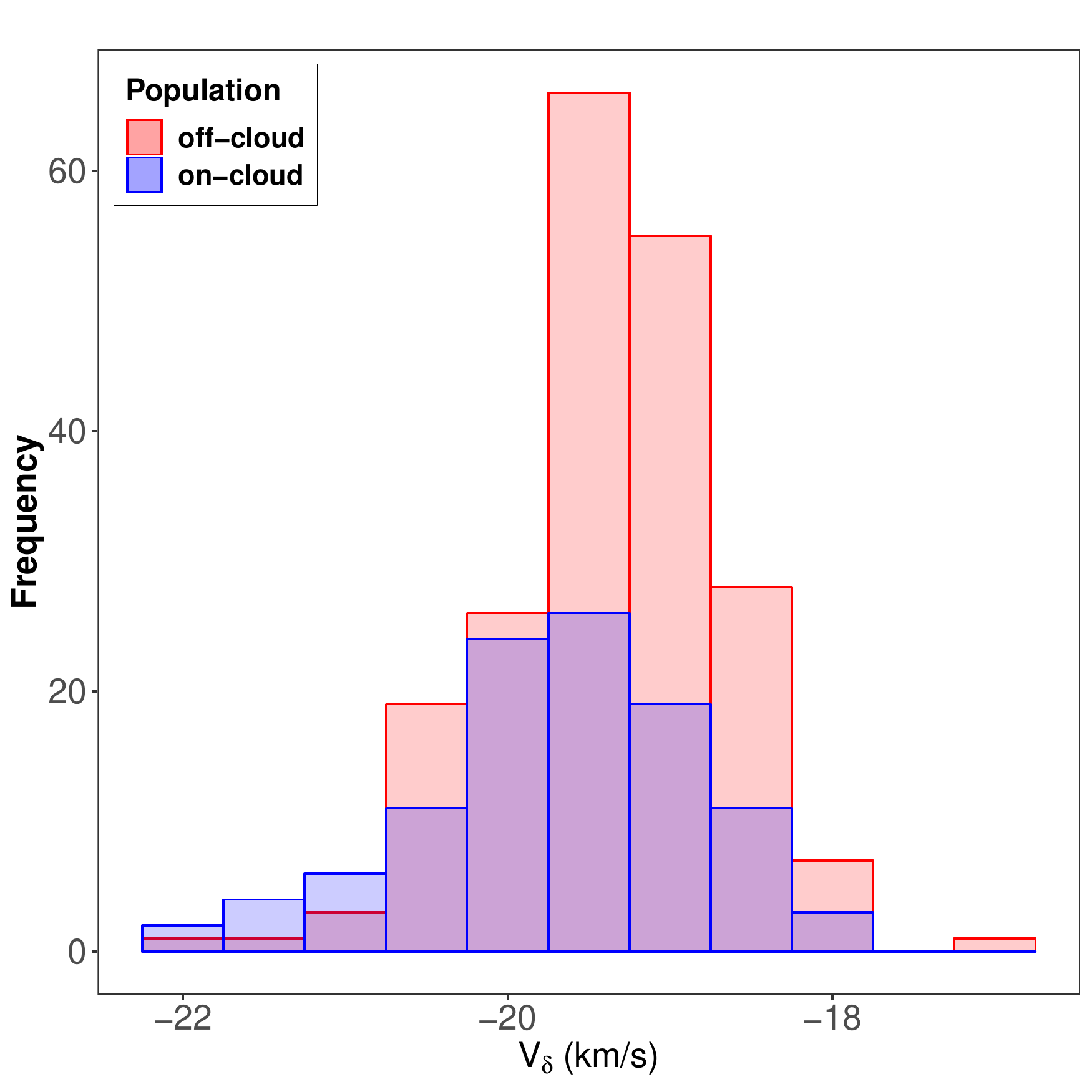}
\caption{
\label{fig_velocity_subgroups}
Distribution of the 2D velocities of Corona-Australis stars in right ascension \textit{(left panel)} and declination \textit{(right panel)}.  
}
\end{center}
\end{figure*}

The discussion about the kinematic properties of the Corona-Australis region in this paper is mostly restricted to the 2D\ tangential velocities of the stars because most members in our sample, in particular the newly discovered off-cloud stars, do not have measured radial velocities in the literature. Figure~\ref{fig_velocity_subgroups} shows the distribution of tangential velocities that we derived from Bayesian inference (as explained above). The existence of two subgroups in our sample is clearly evident once more from the distribution of tangential velocities, in particular, in the component of right ascension. The difference between the mean tangential velocities (in right ascension) of the two subgroups is $2.6\pm0.1$~km/s. In addition, we also verified that the 1D velocity dispersion in each subgroup is about 1~km/s. These results are summarised in Table~\ref{tab_distance}. The typical radial velocity of a few of the stars associated to this region and previously identified in the literature is $V_{r}=-1.1\pm0.5$~km/s \citep{James2006}. Thus, we conclude that the tangential velocity (in declination) is the dominant component in the spatial velocity of Corona-Australis stars.  

\begin{table*}
\centering
\scriptsize{
\caption{Distance and tangential velocity of the subgroups in Corona-Australis.
\label{tab_distance}}
\begin{tabular}{lcccccccc}
\hline\hline
Sample&$N_{stars}$&$d$&\multicolumn{3}{c}{$V_{\alpha}$}&\multicolumn{3}{c}{$V_{\delta}$}\\
&&(pc)&\multicolumn{3}{c}{(km/s)}&\multicolumn{3}{c}{(km/s)}\\
\hline
&&&Mean$\pm$SEM&Median&SD&Mean$\pm$SEM&Median&SD\\
\hline
Off-cloud & 207 & $ 147.9 _{ -0.4 }^{+ 0.3 }$& $ 0.5 \pm 0.1 $& 0.6 & 1.0 & $ -19.4 \pm 0.0 $& -19.3 & 0.7 \\
On-cloud & 106 & $ 152.4 _{ -0.4 }^{+ 0.4 }$& $ 3.1 \pm 0.1 $& 3.1 & 0.6 & $ -19.7 \pm 0.1 $& -19.7 & 0.8 \\
Full sample & 313 & $ 149.4 _{ -0.4 }^{+ 0.4 }$& $ 1.4 \pm 0.1 $& 1.1 & 1.5 & $ -19.5 \pm 0.0 $& -19.4 & 0.8 \\

\hline\hline
\end{tabular}
\tablefoot{We provide for each subgroup the number of stars,  distance derived from the Bayesian approach, mean, standard error of the mean (SEM), median and standard deviation (SD) of the tangential velocity components in right ascension and declination.}
}
\end{table*}

\subsection{Relative ages of the two populations}

In this section, we try to compare the age of the two subgroups by using two proxies: the HR-diagram and the frequency of circumstellar discs. We used the Virtual Observatory SED Analyzer \citep[VOSA,][]{VOSA} to fit the spectral energy distribution (SED) as well as to derive the effective temperature and bolometric luminosities of the stars in our sample. The estimated parameters are used in a subsequent analysis to generate the HR-diagram of the Corona-Australis region. In this context, we used the individual distances derived in Sect.~\ref{section3.2} to fit the SEDs of the stars. The extinction $A_{V}$ is not known for most sources in the sample and we have therefore decided to set it as a free parameter (in the range of 0 mag to 10~mag) to be included in the model fit. We built the SEDs from the Gaia-DR2, 2MASS, and AllWISE photometry provided by ourselves to the VOSA service to avoid erroneous cross-matches when querying these catalogues with the system interface. We cross-matched our sample of stars with the AllWISE catalogue \citep{AllWISE} by using the \texttt{ALLWISE\_BEST\_NEIGHBOUR} table in the Gaia archive and following the same procedure as described in Sect.~\ref{section2.1} for the cross-match with the 2MASS catalogue. Then, we used the BT-Settl \citep{Allard2014} grid of theoretical spectra to fit the SEDs of the stars as well as to derive effective temperatures and bolometric luminosities.

\begin{figure*}
\begin{center}
\includegraphics[width=0.49\textwidth]{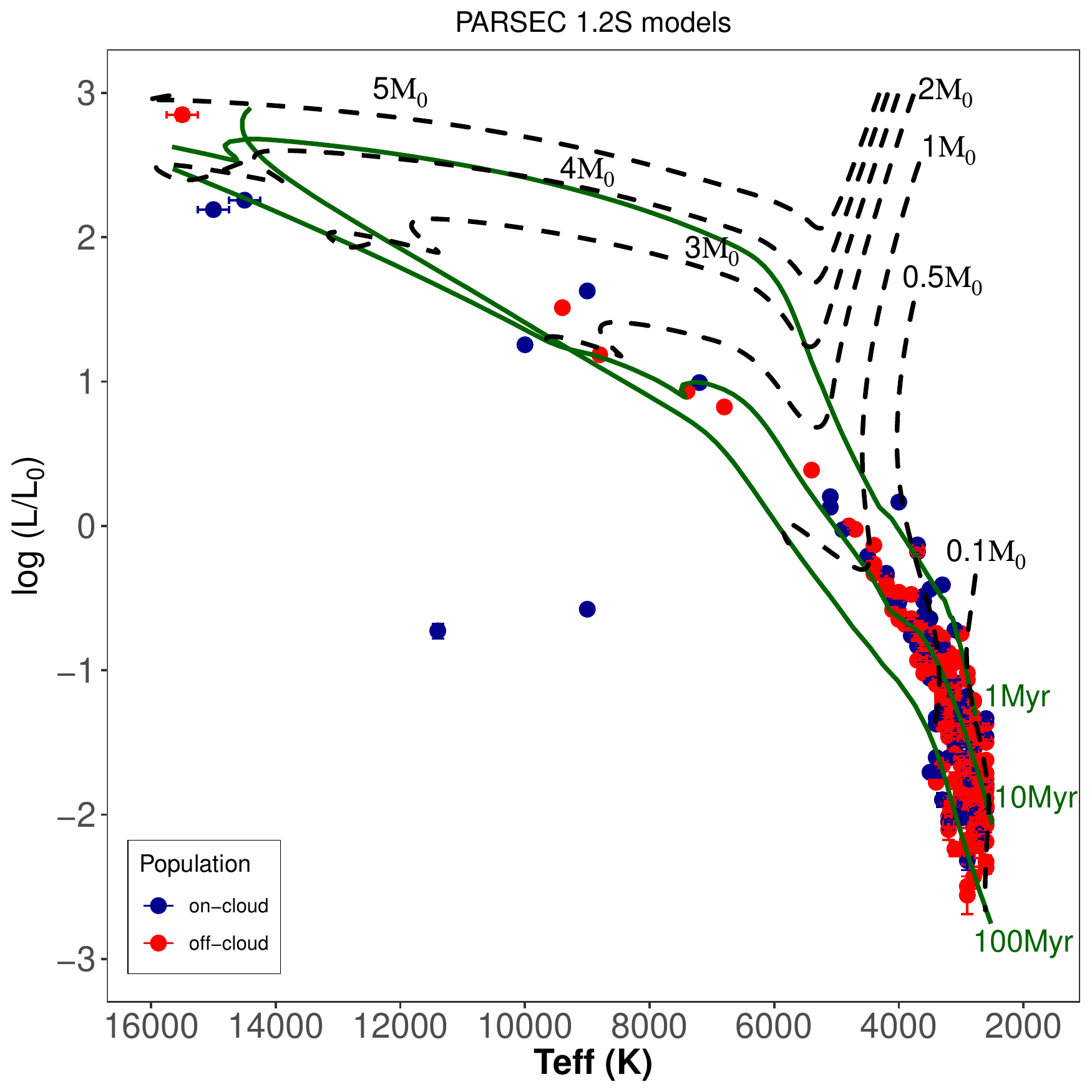}
\includegraphics[width=0.49\textwidth]{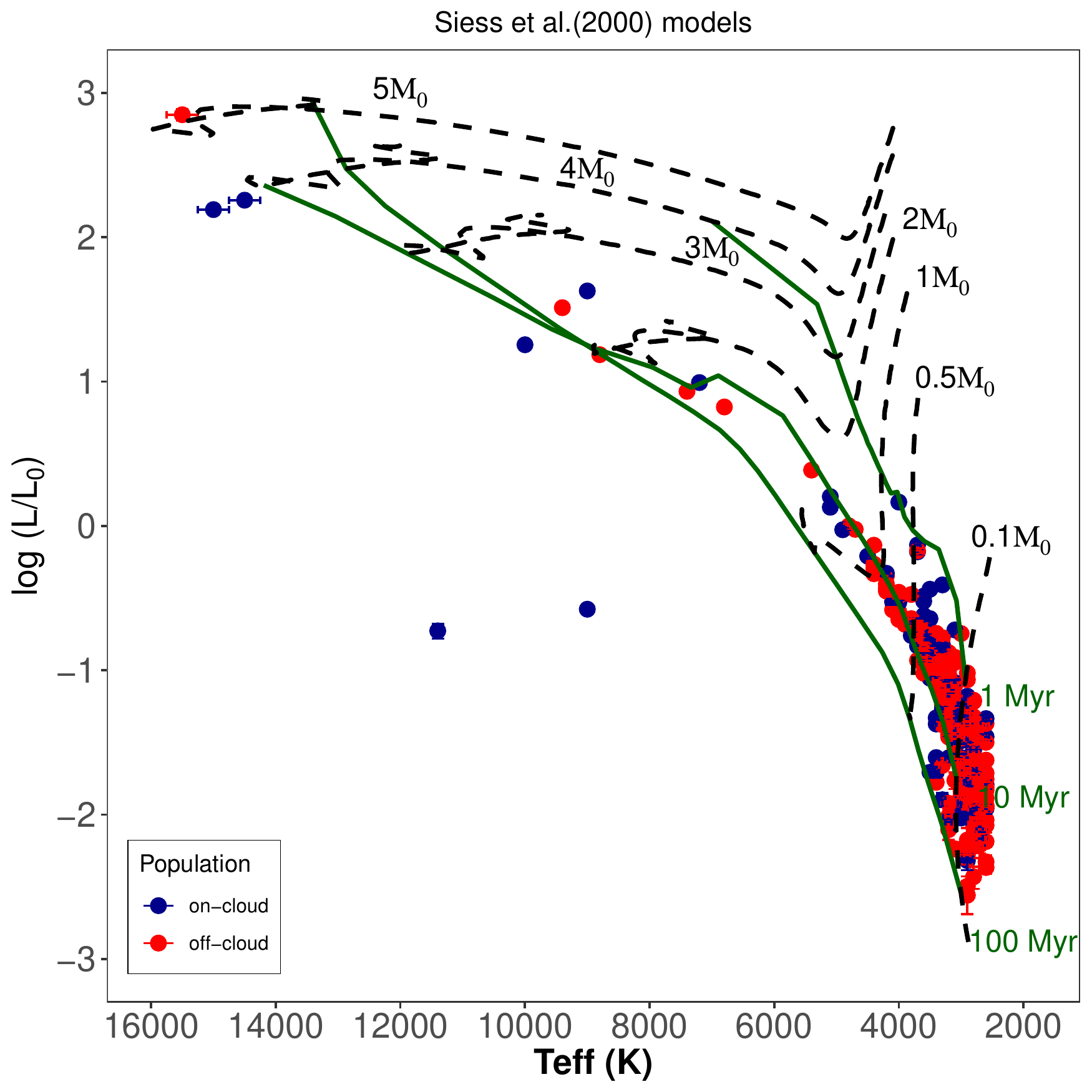}
\includegraphics[width=0.49\textwidth]{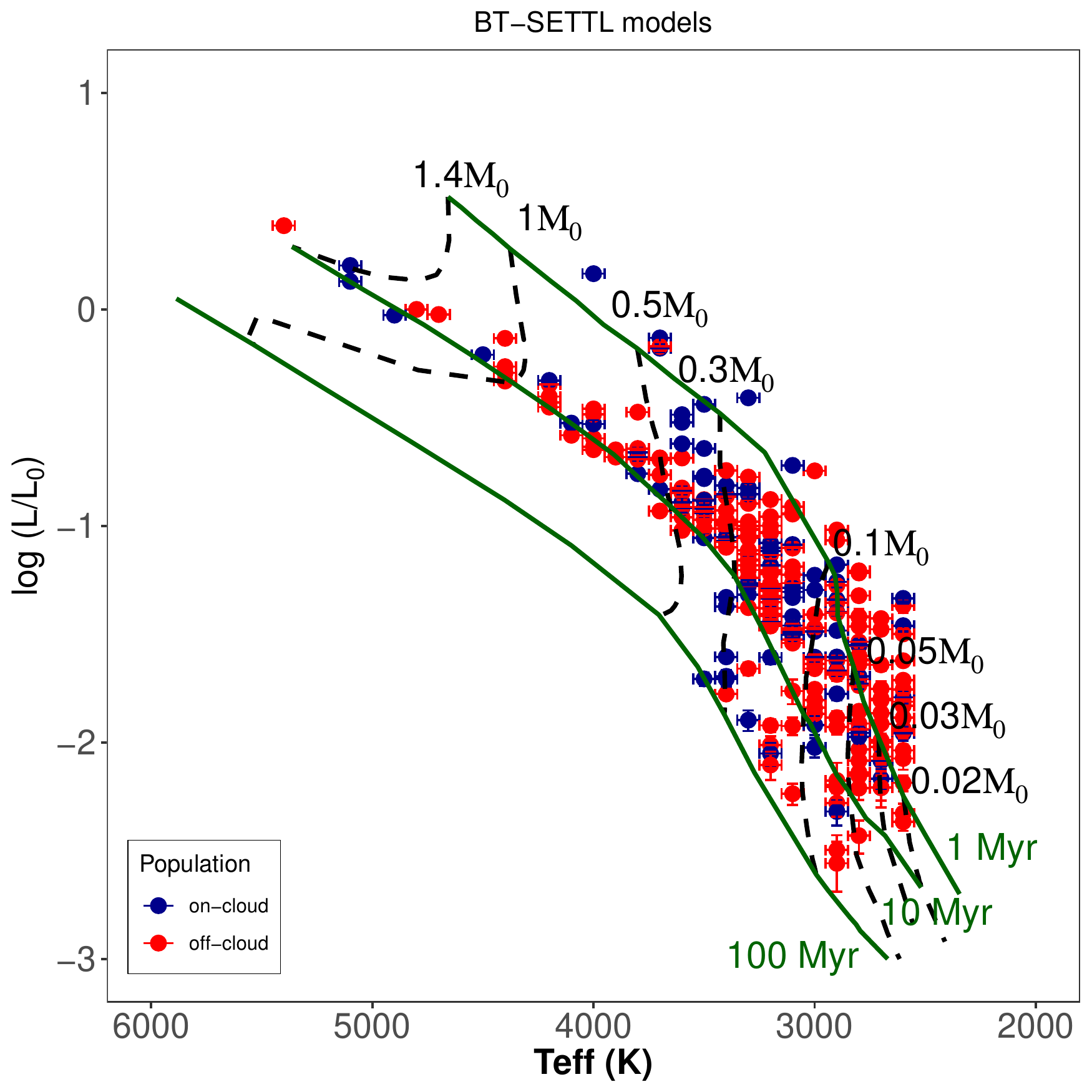}
\includegraphics[width=0.49\textwidth]{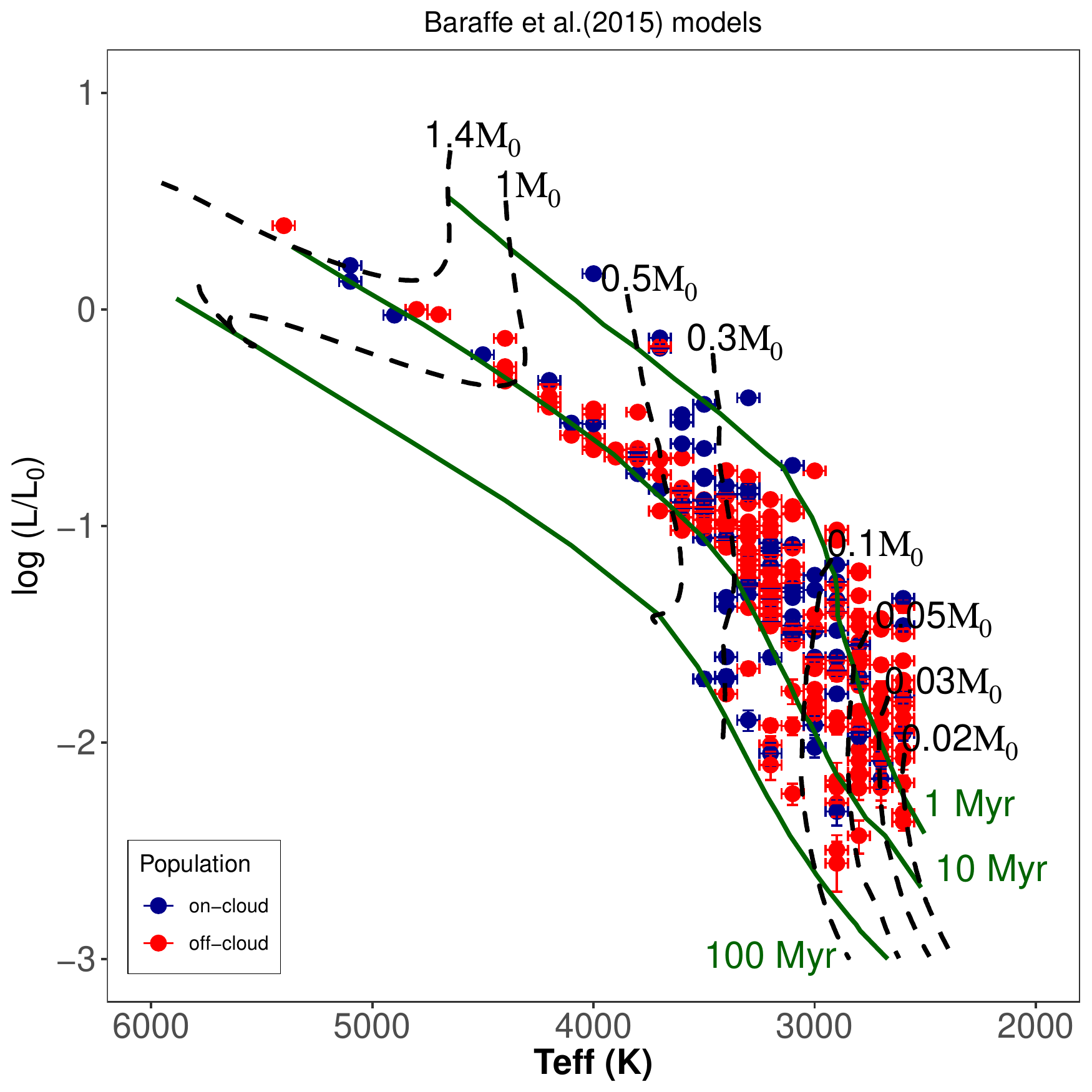}
\caption{
\label{fig_HRD}
HR-diagram of the Corona-Australis region including different evolutionary models (PARSEC~1.2S, \citealt{Siess2000}, BT-Settl and \citealt{BHAC15}). The solid and dashed lines represent isochrones and tracks, respectively. The corresponding ages and masses are indicated in each panel for the various models. The stellar ages and masses are given in units of Myr and solar mass, respectively.
}
\end{center}
\end{figure*}

Figure~\ref{fig_HRD} shows the resulting HR-diagram of our sample including the various evolutionary models for pre-main sequence stars. We used the BT-Settl \citep{Allard2014} and \citet{BHAC15} models to infer the ages and masses of the late-type stars in our sample. For the few sources in our sample that lie outside the region covered by these two models, we used the \citet{Siess2000} and PARSEC~1.2S \citep{Bressan2012} models. It is interesting to note that our sample includes stars with masses ranging from 0.02$M_{\odot}$ to about 5$M_{\odot}$. HD~172910 (Gaia~DR2~6733635914056263296)  is the most massive star identified in our analysis (as anticipated in Sect.~\ref{section2.4}), but our mass estimate is still smaller than the value of $M=7.2\pm 0.2M_{\odot,}$ which was previously derived by \citet{Tetzlaff2011}. The discrepancy between the two studies can be explained by the different data (e.g. parallax and spectral type) used in each case to derive the stellar parameters (e.g. luminosity and effective temperature) and estimate the stellar mass from evolutionary models. We note that most sources in our sample are younger than 10~Myr, and a number of them also appear to be younger than 1~Myr. Of course, some of these sources (above the 1~Myr isochrone) could also be binaries or high-order multiple systems, but this hypothesis requires further investigation with follow-up observations. The median age of the sample inferred from the 218 sources in the area covered by the BT-Settl isochrones is 6~Myr. When we compare the on-cloud and off-cloud populations in our sample, we find the median ages of 5 Myr and 6~Myr, respectively. This suggests that the on-cloud population is somewhat younger and the small difference between these age estimates confirms that the two populations are indeed part of the same star-forming region.   

Let us now compare the disc properties of the two populations to search for any additional hints of evolution. Circumstellar discs are indeed known to evolve and disappear relatively rapidly within the first 10~Myr \citep[e.g.][]{2014A&A...561A..54R}. The occurrence of circumstellar discs in a group of young stars can therefore provide some hints about the evolutionary status of the group, if not in an absolute way, at least in a relative way. \citet{Koenig2014} developed a classification scheme based on 2MASS and AllWISE photometry that we use here to classify the stars in our sample. This method uses colours and magnitudes to define the locus of Class~I, Class~II, and transition disc \footnote{Transition discs are defined as discs with inner opacity holes and reduced levels of near- and mid-infrared excess emission, which represent an intermediate stage between classical T~Tauri and weak-line T~Tauri stars \citep{Cieza2012}.} objects in a number of colour-colour diagrams depending on the presence or absence of infrared excess emission of the sources. The method also identifies a number of astrophysical objects e.g. asymptotic giant branch (AGB) stars, classical Be stars, star-forming galaxies, and active galactic nucleus (AGN) which have been frequently misclassified as YSOs in the past \citep[see e.g.][]{Vieira2011}. However, given the very young ages and distances that we derived in this study for the Corona-Australis stars, we can discard the existence of such contaminants in our sample. Thus, we proceed as follows to classify our YSOs. 

\begin{figure*}
\begin{center}
\includegraphics[width=0.49\textwidth]{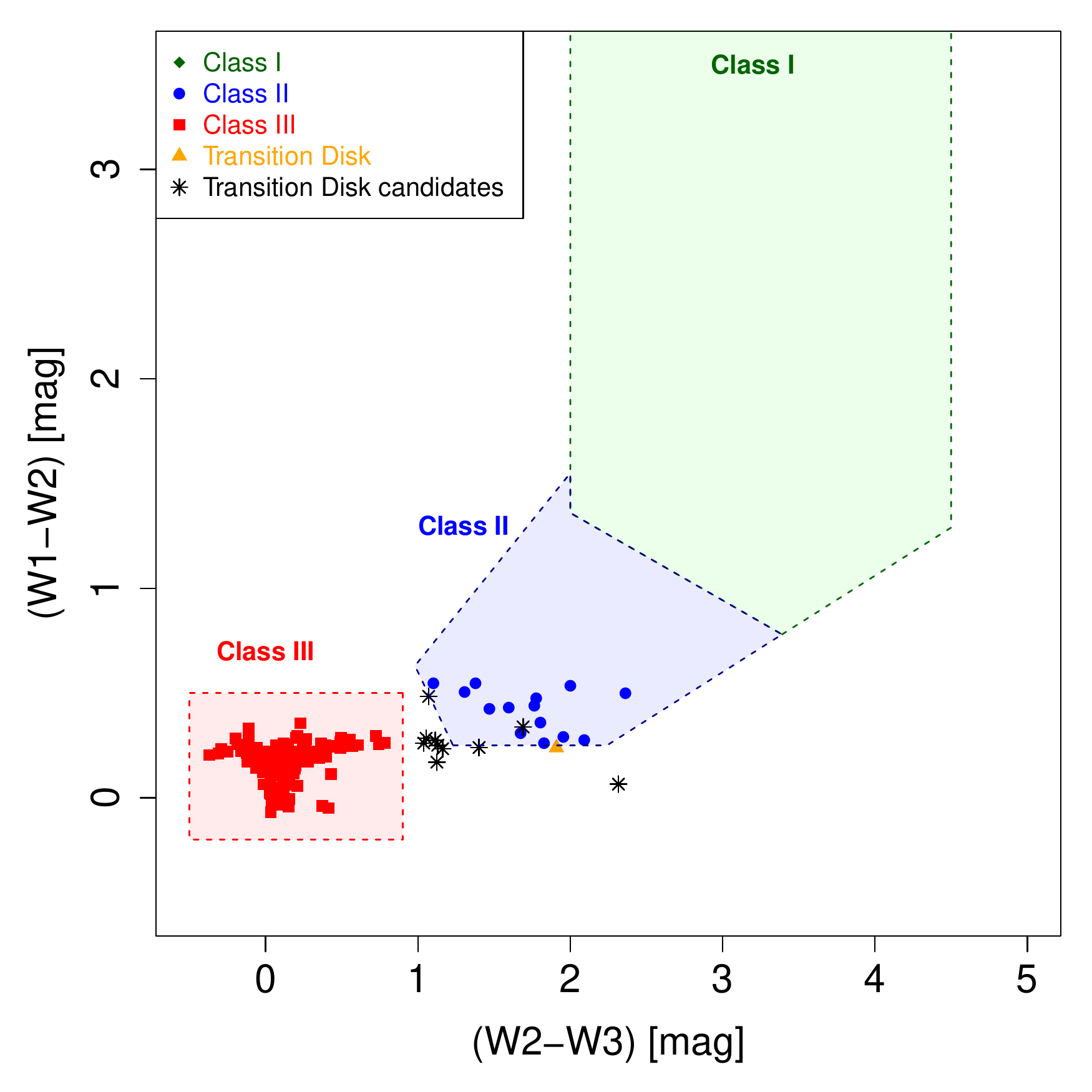}
\includegraphics[width=0.49\textwidth]{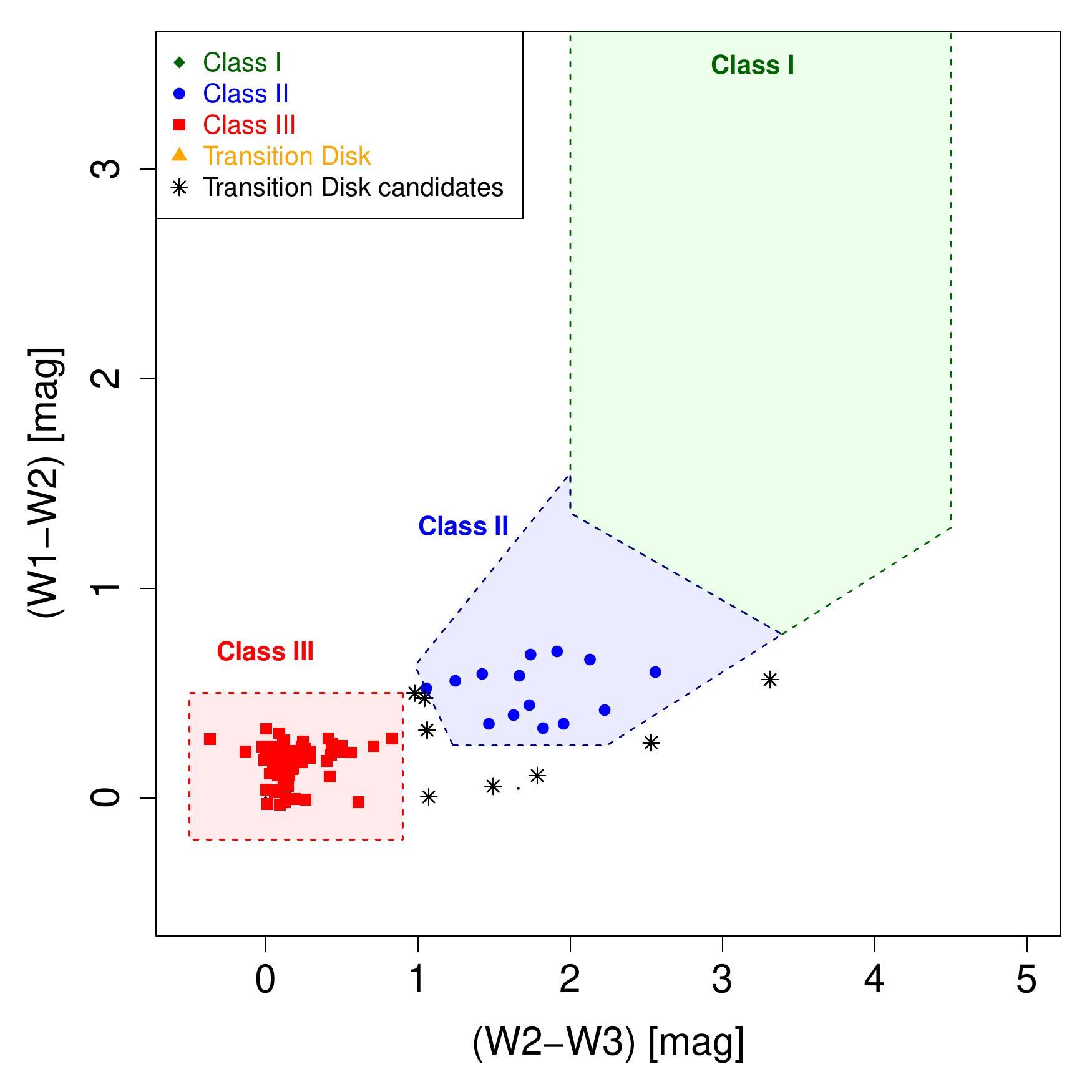}
\caption{
\label{fig_Koenig2014}
Colour-colour diagram used in the YSO classification scheme by \citet{Koenig2014} applied to the off-cloud \textit{(left panel)} and on-cloud \textit{(right panel)} populations of stars in Corona-Australis. 
}
\end{center}
\end{figure*}

\bigskip
We applied the photometric selection criteria described in Sect.~3.2 of \citet{Koenig2014} to mitigate fake source contamination in the AllWISE catalogue. This reduced the sample to 262 stars. Then, we applied the YSO classification scheme to the remaining stars and classified them into Class~I, Class~II, and transition disc stars. Figure~\ref{fig_Koenig2014} illustrates, as an example, one of the colour-colour diagrams used by the classification scheme. We note that most stars in the sample fall between $W2-W3 < 1.0$ and $W1-W2 < 0.5,$ which also coincides with the region where both Class~III and AGB stars reside \citep[see Fig.~5 of ][]{Koenig2014}. As explained before, we do not expect our sample to be contaminated by AGB stars and we have therefore classified these sources as Class~III stars. We also note the existence of a number of sources (marked with black asterisks in Figure~\ref{fig_Koenig2014}) with significant infrared excess that fall beyond the Class~II locus. As shown in Figure~5 of \citet{Koenig2014}, this region of the diagram is also populated by transition disc objects, which still exhibit important infrared excess emission as well as edge-on discs. For the moment, we have classified these stars as new transition disc candidates, but this requires confirmation and they are listed in Table~\ref{tab_TD}. As shown in this figure, only one star (namely, Gaia~DR2~6733045308825699328) has been directly classified as a transition disc object by the \citet{Koenig2014} classification scheme. We did not detect any Class~I stars in our sample, although such sources are known to exist in the Corona-Australis region  (as explained in Sect.~\ref{section1}). Such deeply embedded sources are indeed not expected to be detected by the optical sensors of the Gaia satellite and we verified that all Class~0/I sources of the Coronet cluster listed in Table~2 of \citet{Neuhauser2008} were not included in our membership analysis because they do not have Gaia-DR2 data.  

Table~\ref{tab_YSO} summarises the results of this classification for the two populations of stars in Corona-Australis. Interestingly, the frequency of Class~II stars harboring circumstellar material is higher by a factor of almost two for the on-cloud population, suggesting that the on-cloud population is younger than its off-cloud counterpart. Altogether, this suggests that the more dispersed off-cloud stars form an older, that is, more evolved, population of YSOs. 

\begin{table*}
\centering
\scriptsize{
\caption{Transition disk stars and candidates in the sample of Corona-Australis members.
\label{tab_TD}}
\begin{tabular}{lccccccccc}
\hline\hline
Source Identifier&$\alpha$&$\delta$&$J$&$H$&$K$&$W1$&$W2$&$W3$&$W4$\\
&(h:m:s) &($^{\circ}$ $^\prime$ $^\prime$$^\prime$)&(mag)&(mag)&(mag)&(mag)&(mag)&(mag)&(mag)\\
\hline\hline
Gaia DR2 6734929047140217216 & 18 29 27.99 & -34 21 51.1 & 14.350 & 13.871 & 13.430 & 13.200 & 12.926 & 11.814 & 8.746 \\
Gaia DR2 6733045308825699328 & 18 35 21.41 & -37 34 18.8 & 11.653 & 10.974 & 10.708 & 10.404 & 10.165 & 8.256 & 5.348 \\
Gaia DR2 6736763921557552000 & 18 38 51.90 & -32  06 17.4 & 15.362 & 14.886 & 14.615 & 14.431 & 14.366 & 12.050 & 8.556 \\
Gaia DR2 6733973361050980352 & 18 39 12.35 & -34 52 14.6 & 15.209 & 14.540 & 14.146 & 13.749 & 13.411 & 11.721 & 8.825 \\
Gaia DR2 6736317421080943232 & 18 39 32.60 & -33 32 25.5 & 14.696 & 14.067 & 13.646 & 13.434 & 13.264 & 12.140 & 8.523 \\
Gaia DR2 6729972895366111232 & 18 42  05.08 & -37 55 26.4 & 14.647 & 13.977 & 13.591 & 13.034 & 12.550 & 11.480 & 7.992 \\
Gaia DR2 6735207738950579456 & 18 43 49.49 & -35  06 17.0 & 14.575 & 13.867 & 13.435 & 13.072 & 12.788 & 11.735 & 8.372 \\
Gaia DR2 6730572812103102848 & 18 43 49.88 & -36 28 17.1 & 14.219 & 13.599 & 13.312 & 13.072 & 12.838 & 11.676 & 8.849 \\
Gaia DR2 6730394553778846720 & 18 44 41.47 & -37 12 19.6 & 14.514 & 13.835 & 13.516 & 13.289 & 13.038 & 11.908 & 8.674 \\
Gaia DR2 6735689977887062144 & 18 45 16.66 & -33  00 45.1 & 14.957 & 14.330 & 13.906 & 13.795 & 13.534 & 12.496 & 8.993 \\
Gaia DR2 6730265223722584320 & 18 48 18.29 & -37 11  03.8 & 13.308 & 12.089 & 11.437 & 10.779 & 10.279 & 9.299 & 7.673 \\
Gaia DR2 6730252167025513984 & 18 50  08.57 & -37 15 34.2 & 12.718 & 12.165 & 11.851 & 11.698 & 11.458 & 10.057 & 7.785 \\
Gaia DR2 6731011315385156224 & 18 52 17.31 & -37  00 12.4 & 9.772 & 9.141 & 9.007 & 8.837 & 8.731 & 6.948 & 3.109 \\
Gaia DR2 6731902087299776000 & 18 52 52.07 & -36 14 23.8 & 14.774 & 14.005 & 13.550 & 13.273 & 12.950 & 11.891 & 8.845 \\
Gaia DR2 6730829994739866624 & 18 59 50.95 & -37  06 31.9 & 13.983 & 13.104 & 12.565 & 12.198 & 11.722 & 10.678 & 8.615 \\
Gaia DR2 6731216408658348032 & 19  00 58.05 & -36 45  05.3 & 10.383 & 9.512 & 9.196 & 8.957 & 8.902 & 7.409 & 4.001 \\
Gaia DR2 6730822302462396160 & 19  01  03.26 & -37  03 39.7 & 6.719 & 6.778 & 6.740 & 6.701 & 6.697 & 5.626 & 3.517 \\
Gaia DR2 6730822023280685440 & 19  01 20.85 & -37  03  03.2 & 13.233 & 12.687 & 12.402 & 12.196 & 11.933 & 9.402 & 6.780 \\
Gaia DR2 6731197442076732928 & 19  01 33.58 & -37  00 30.9 & 15.178 & 14.526 & 13.972 & 13.535 & 12.971 & 9.660 & 6.971 \\

\hline\hline
\end{tabular}
\tablefoot{We provide for each source the Gaia-DR2 identifier and position, infrared photometry from the 2MASS and  AllWISE catalogues.}
}
\end{table*}

\begin{table}
\centering
\scriptsize{
\caption{Number of stars classified in the various YSO subclasses based on the sample of 262 sources with available AllWISE photometry.
\label{tab_YSO}}
\begin{tabular}{lcccc}
\hline\hline
Sample&Class~I&Class~II&Class~III&Transition Disk\\
&&&&\\
\hline\hline
Off-cloud&0 (0\%)&14 (8.2\%)&146 (85.4\%)&11 (6.4\%)\\
On-cloud&0 (0\%)&14 (15.4\%)&69 (75.8\%) &8 (8.8\%)\\
\hline
Full Sample&0 (0\%)&28 (10.7\%)&215 (82.0\%)&19 (7.3\%)\\
\hline\hline
\end{tabular}
\tablefoot{In the parenthesis, we provide the relative fraction of the various subclasses for each sample.}}
\end{table}

\subsection{Spatial distribution of Corona-Australis stars}

\begin{figure*}
\begin{center}
\includegraphics[width=0.33\textwidth]{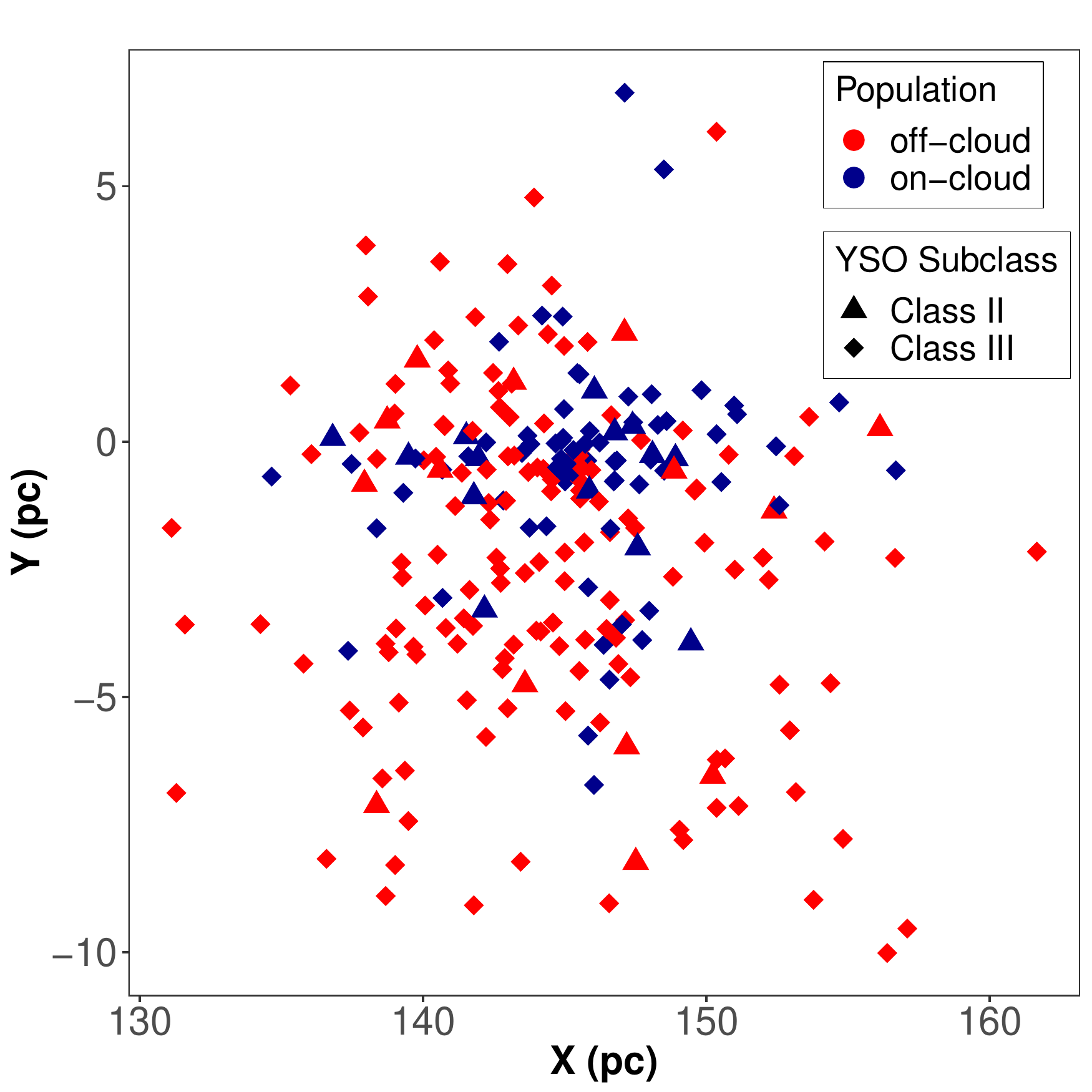}
\includegraphics[width=0.33\textwidth]{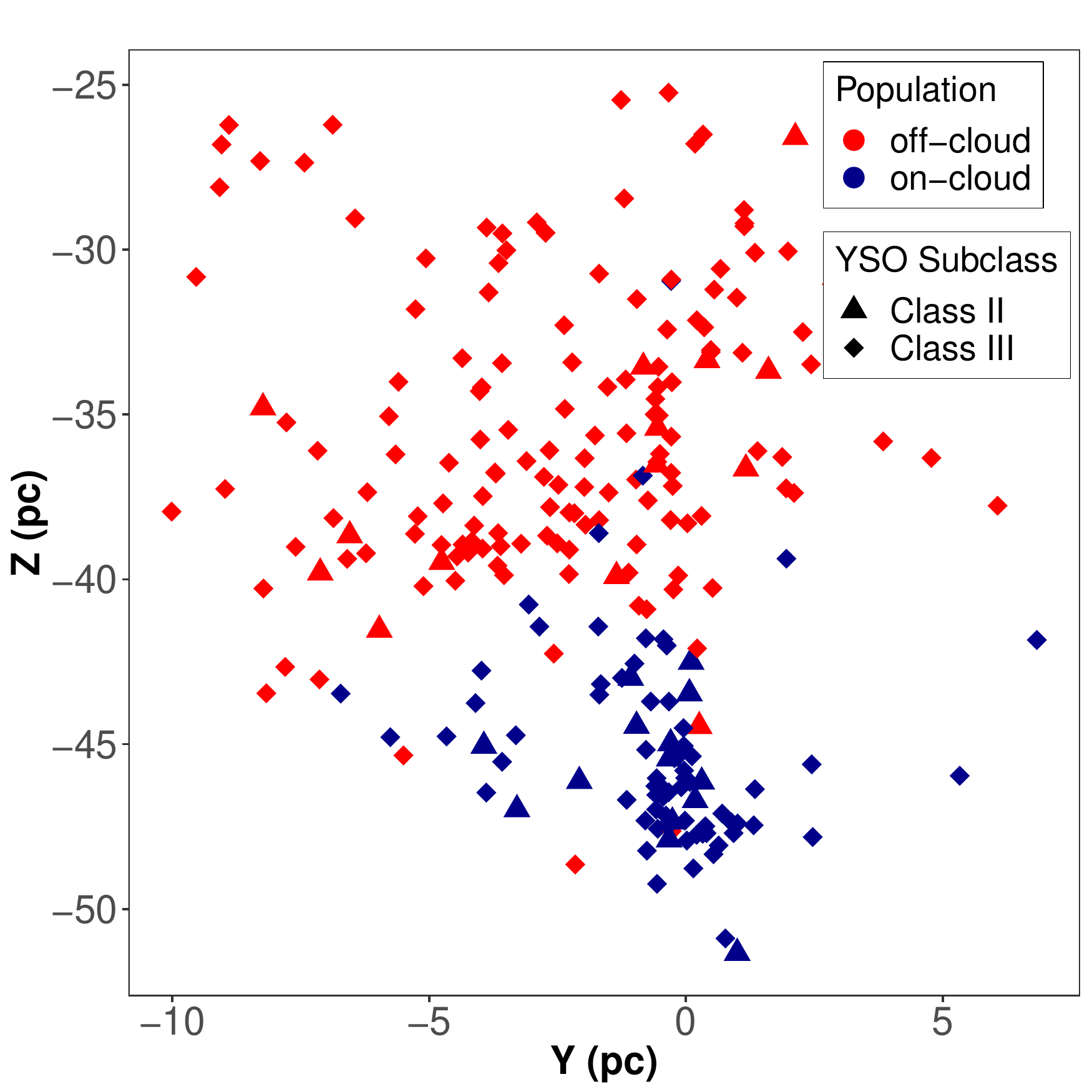}
\includegraphics[width=0.33\textwidth]{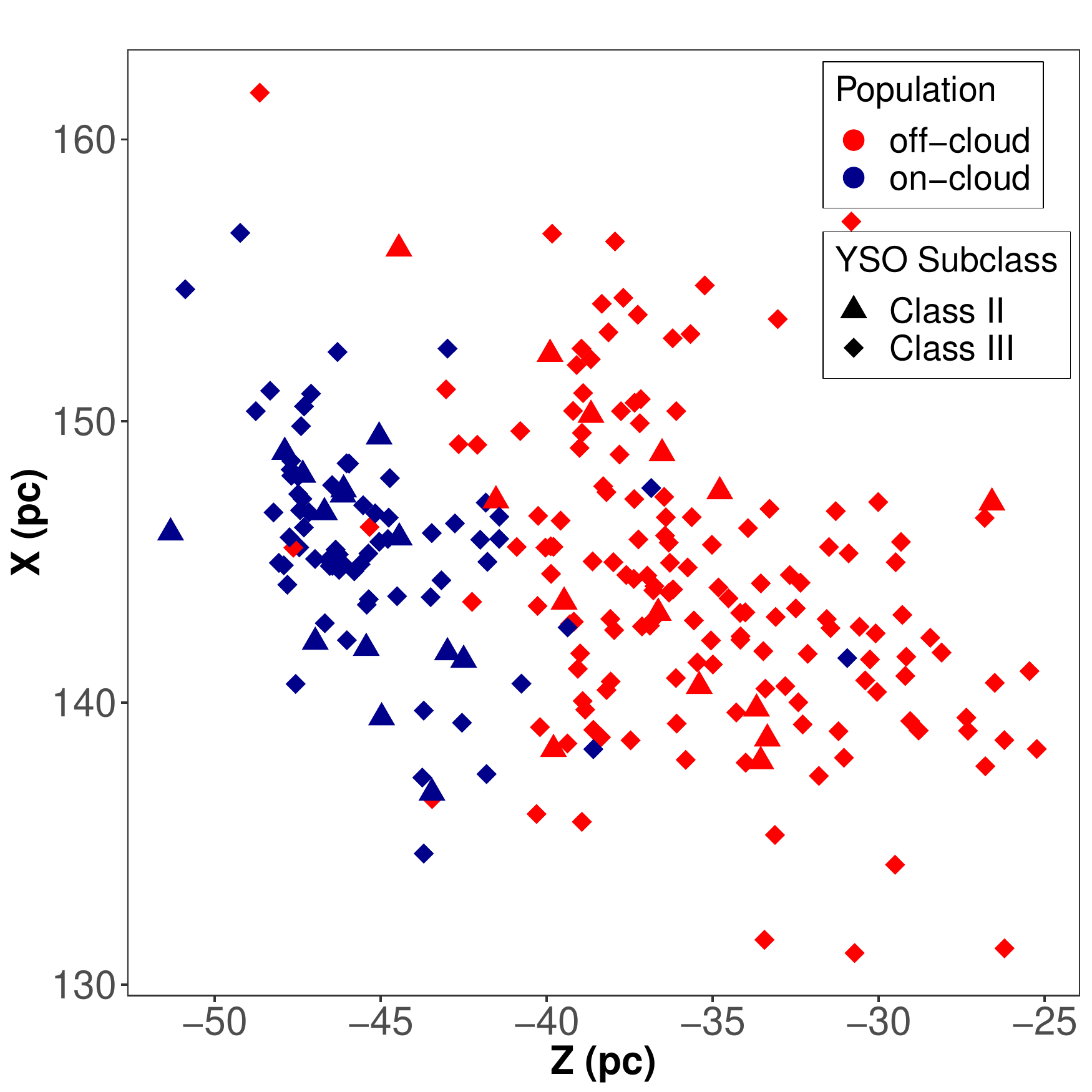}
\caption{
\label{fig_XYZ}
3D spatial distribution of the YSOs (and their subclasses) in Corona-Australis. 
}
\end{center}
\end{figure*}

The 3D spatial distribution of the YSOs, and the various subclasses, in the two populations are illustrated in Figure~\ref{fig_XYZ}. It is apparent that the two subgroups of stars are located at different positions with respect to the Galactic plane. The median distance of the on-cloud and off-cloud populations to the Galactic plane are $-46$~pc and $-36$~pc, respectively. In addition, we observe that the Class~II stars in the on-cloud population are more clustered in space as compared to the off-cloud population.

This scenario of overlapping younger and older populations of YSOs is also observed in other nearby star-forming regions. For example, \citet{Galli2013} show that the on-cloud and off-cloud populations of YSOs in the Lupus region exhibit different kinematic properties. \citet{Galli2015} confirm that the off-cloud stars, which were mostly weak-line T~Tauri stars (i.e. Class~III stars), are indeed older than the on-cloud stars in that region. \citet{Lopez-Marti2013} identify a number of discless stars in the Chamaeleon star-forming region that tend to be located in the outskirts of the dark clouds, which host most of the known YSOs in this region \citep[see e.g.][]{Luhman2004,Luhman2007}. \citet{Kraus2017} and \citet{Zhang2018} also report on a distributed population of young stars in the Taurus region, which is older ($>10$~Myr) than the classical members of the region \citep{Luhman2018}.  Another well-known example is the Orion complex, which is made up of several groups and clusters of YSOs with different ages \citep[see][]{Alves2012,Bouy2014,Kounkel2018,Zari2019,Chen2019}. Our analysis conducted in this paper shows that Corona-Australis is one more such substructured star-forming region that will require further investigation to understand its star formation history. 

One interesting point while comparing Corona-Australis with Lupus, for example, is that the sample of on-cloud stars in the latter is at least twice as large when compared to the off-cloud population \citep[see e.g. Table~2 of][]{Galli2015}. This contrasts with the results that we obtain here for the Corona-Australis region (see e.g. Table~\ref{tab_subgroups}) where the off-cloud stars clearly dominate our sample of cluster members. As mentioned before, some of the known YSOs, which were  previously identified in the literature (e.g. the deeply embedded Class~I stars), are not discussed here because they are not included in the Gaia-DR2 catalogue.  In addition, we also applied a conservative approach based on the RUWE selection criterion to filter the sources with reliable Gaia-DR2 data for the membership analysis (as explained in Sect.~\ref{section2.1}). Although our new sample of cluster members significantly improves the current census of stars in this region, we argue that our list is not complete yet. Our team is currently refurbishing the methodology developed by \citet{Olivares2018}, which was based on hierarchical Bayesian models, to perform membership analysis in regions of high extinction, and we will soon be able to provide a more complete census of the stars in the densest cores of Corona-Australis. 

\section{Conclusions}\label{section4}

We applied a probabilistic method based on Gaia-DR2 data to infer membership probabilities of more than $10^{7}$ sources over 128~deg$^{2}$ in the Corona-Australis star-forming region. We identified 313 stars that are probable members of the young association of stars in this region. We confirm 51 stars with available Gaia-DR2 data, which have been previously identified in the literature and detected 262 new members. This result increases the number of confirmed cluster members (with available Gaia-DR2 astrometry) in this region by a factor of almost 5. 

Our analysis reveals the existence of a distributed population of stars beyond the densest cores, which is  located in the northern region of the dark cloud complex. This off-cloud population is almost twice as large, in terms of the number of stars, as the on-cloud population, which is more concentrated in the region of the main molecular clouds. The most discriminant features between the two populations in our sample are the proper motion and tangential velocity in right ascension. The distance variation along the line of sight between the two subgroups is $4.5\pm0.1$~pc. We derived the distance of $149.4^{+0.4}_{-0.4}$~pc to Corona-Australis based on Bayesian inference, which exceeds previous estimates by about 20~pc. The HR-diagram that we obtain in this study shows that the stars selected in our membership analysis are mostly younger than 10~Myr, which unambiguously confirms them to be YSOs. The stellar masses range from about 0.02$M_{\odot}$ to $5M_{\odot}$, and the median ages of the on-cloud and off-cloud populations are 5 Myr and 6~Myr, respectively. We classify 28~YSOs as Class~II stars, 215~YSOs as Class~III stars, and 19~YSOs as transition disc objects (and candidates) based on their infrared excess emission derived from AllWISE photometry. We report that the frequency of accretors, that is, Class~II stars, is twice as large for the on-cloud population and this subgroup hosts the youngest stars in our sample. Altogether, this suggests that the off-cloud stars form a more evolved population of YSOs in the Corona-Australis region, as is observed in other nearby star-forming regions.  

This study significantly increases the number of known YSOs in Corona-Australis, but the census of the stellar (and substellar) content in this region is still not complete yet. We restricted our analysis to the Gaia-DR2 data, which are of limited use in the region of the densest cores with high extinction. We are currently measuring the proper motions of faint sources based on archival images, and our own observations, as part of the DANCe project \citep{Bouy2013} to complement the Gaia-DR2 catalogue in this region and to extend upon the methodology developed by \citet{Olivares2018} in order to perform membership analysis in regions of high (and variable) extinction. In addition, we are also starting an observing campaign to characterise the newly discovered YSOs in this study. We will present the results of these analyses and derive the initial mass function of the Corona-Australis association in a companion paper.

\begin{acknowledgements}
We thank the referee for constructive comments that helped us to improve the manuscript.
This research has received funding from the European Research Council (ERC) under the European Union’s Horizon 2020 research 
and innovation programme (grant agreement No 682903, P.I. H. Bouy), and from the French State in the framework of the 
``Investments for the future” Program, IdEx Bordeaux, reference ANR-10-IDEX-03-02. This research has made use of the SIMBAD database,
operated at CDS, Strasbourg, France. This work has made use of data from the European Space Agency (ESA) mission {\it Gaia} (\url{https://www.cosmos.esa.int/gaia}), processed by the {\it Gaia} Data Processing and Analysis Consortium (DPAC, \url{https://www.cosmos.esa.int/web/gaia/dpac/consortium}). Funding for the DPAC has been provided by national institutions, in particular the institutions participating in the {\it Gaia} Multilateral Agreement. This publication makes use of data products from the Wide-field Infrared Survey Explorer, which is a joint project of the University of California, Los Angeles, and the Jet Propulsion Laboratory/California Institute of Technology, funded by the National Aeronautics and Space Administration. This publication makes use of VOSA, developed under the Spanish Virtual Observatory project supported by the Spanish MINECO through grant AyA2017-84089.
\end{acknowledgements}

\bibliographystyle{aa} 
\bibliography{references} 

\begin{appendix}
\section{Tables (online material)}\label{appendix_tables}

\begin{landscape}
\begin{table}
\caption{Properties of the 313 cluster members selected from our membership analysis in Corona-Australis. (This table will be available in its entirety in machine-readable form.) }\label{tab_members}
\scriptsize{
\begin{tabular}{lcccccccccll}
\hline\hline
Source Identifier&$\alpha$&$\delta$&$\mu_{\alpha}\cos\delta$&$\mu_{\delta}$&$\varpi$&Probability&$d$&$V_{\alpha}$&$V_{\delta}$&Population&Subclass\\
&(h:m:s) &($^{\circ}$ $^\prime$ $^\prime$$^\prime$)&(mas/yr)&(mas/yr)&(mas)&&(pc)&(km/s)&(km/s)&&\\
\hline\hline
Gaia DR2 6728162648254144128 & 18 20 42.48 & -37  01 41.6 & $ -2.567 \pm 0.256 $& $ -26.763 \pm 0.196 $& $ 6.679 \pm 0.161 $& $ 0.9914 $& $ 149.3 ^{+ 4.6 }_{ -3.7 } $& $ -1.8 ^{+ 0.2 }_{ -0.3 } $& $ -19.0 ^{+ 0.6 }_{ -0.7 } $& Off-cloud & Class~III \\
Gaia DR2 6728140520578676480 & 18 21 55.13 & -37 18 05.4 & $ -3.481 \pm 0.097 $& $ -29.498 \pm 0.079 $& $ 7.045 \pm 0.065 $& $ 0.9852 $& $ 141.4 ^{+ 2.4 }_{ -2.2 } $& $ -2.3 ^{+ 0.1 }_{ -0.1 } $& $ -19.8 ^{+ 0.4 }_{ -0.4 } $& Off-cloud & Class~III \\
Gaia DR2 6728030230146525568 & 18 24 08.86 & -37 18 05.1 & $ -2.223 \pm 0.127 $& $ -26.808 \pm 0.109 $& $ 6.210 \pm 0.070 $& $ 0.9792 $& $ 160.4 ^{+ 3.5 }_{ -2.7 } $& $ -1.7 ^{+ 0.1 }_{ -0.2 } $& $ -20.5 ^{+ 0.4 }_{ -0.5 } $& Off-cloud & Class~III \\
Gaia DR2 6727979343370860288 & 18 24 13.57 & -37 30 58.7 & $ -1.700 \pm 0.144 $& $ -27.897 \pm 0.125 $& $ 6.880 \pm 0.089 $& $ 0.9999 $& $ 144.8 ^{+ 2.8 }_{ -2.7 } $& $ -1.2 ^{+ 0.1 }_{ -0.2 } $& $ -19.2 ^{+ 0.4 }_{ -0.4 } $& Off-cloud & Class~III \\
Gaia DR2 6728028404761019520 & 18 24 19.15 & -37 24 08.6 & $ -1.559 \pm 0.464 $& $ -28.330 \pm 0.406 $& $ 6.497 \pm 0.285 $& $ 0.9951 $& $ 153.8 ^{+ 8.1 }_{ -6.1 } $& $ -1.1 ^{+ 0.4 }_{ -0.4 } $& $ -20.8 ^{+ 1.1 }_{ -1.1 } $& Off-cloud & \nodata \\
Gaia DR2 6728301775168256128 & 18 24 19.86 & -36 14 28.4 & $ -1.835 \pm 0.107 $& $ -29.759 \pm 0.086 $& $ 7.192 \pm 0.080 $& $ 0.9933 $& $ 138.6 ^{+ 2.4 }_{ -2.2 } $& $ -1.2 ^{+ 0.1 }_{ -0.1 } $& $ -19.5 ^{+ 0.3 }_{ -0.4 } $& Off-cloud & \nodata \\
Gaia DR2 6728036277460482176 & 18 24 20.42 & -37 15 14.2 & $ -1.416 \pm 0.064 $& $ -28.218 \pm 0.053 $& $ 7.020 \pm 0.035 $& $ 0.9904 $& $ 141.9 ^{+ 2.2 }_{ -2.1 } $& $ -1.0 ^{+ 0.1 }_{ -0.1 } $& $ -19.0 ^{+ 0.3 }_{ -0.3 } $& Off-cloud & Class~III \\
Gaia DR2 6728067471780966016 & 18 25 03.48 & -36 55 35.4 & $ -2.293 \pm 0.224 $& $ -27.370 \pm 0.203 $& $ 7.004 \pm 0.142 $& $ 0.9998 $& $ 142.3 ^{+ 4.0 }_{ -3.3 } $& $ -1.5 ^{+ 0.2 }_{ -0.2 } $& $ -18.5 ^{+ 0.5 }_{ -0.5 } $& Off-cloud & Class~III \\
Gaia DR2 6728069984365602944 & 18 26 01.92 & -36 57 57.1 & $ -1.815 \pm 0.071 $& $ -30.899 \pm 0.066 $& $ 7.433 \pm 0.040 $& $ 0.9738 $& $ 134.1 ^{+ 2.0 }_{ -1.9 } $& $ -1.2 ^{+ 0.1 }_{ -0.1 } $& $ -19.6 ^{+ 0.3 }_{ -0.3 } $& Off-cloud & Class~III \\
Gaia DR2 4044416569172725888 & 18 26 30.93 & -34 18 00.0 & $ -2.429 \pm 0.552 $& $ -27.507 \pm 0.500 $& $ 6.952 \pm 0.260 $& $ 0.9916 $& $ 143.7 ^{+ 6.5 }_{ -4.9 } $& $ -1.7 ^{+ 0.4 }_{ -0.5 } $& $ -18.9 ^{+ 0.8 }_{ -0.9 } $& Off-cloud & \nodata \\
Gaia DR2 4044416328691645824 & 18 26 36.53 & -34 18 32.7 & $ -1.008 \pm 0.247 $& $ -26.848 \pm 0.223 $& $ 6.949 \pm 0.101 $& $ 0.9997 $& $ 143.4 ^{+ 3.0 }_{ -2.8 } $& $ -0.7 ^{+ 0.2 }_{ -0.2 } $& $ -18.3 ^{+ 0.4 }_{ -0.4 } $& Off-cloud & Class~III \\
Gaia DR2 4044819093492085120 & 18 27 07.60 & -33 51 13.9 & $ -2.527 \pm 0.426 $& $ -28.243 \pm 0.380 $& $ 6.766 \pm 0.254 $& $ 0.9958 $& $ 147.6 ^{+ 7.0 }_{ -5.3 } $& $ -1.8 ^{+ 0.4 }_{ -0.4 } $& $ -20.1 ^{+ 0.9 }_{ -0.9 } $& Off-cloud & \nodata \\
Gaia DR2 6734509613481584896 & 18 27 28.86 & -35  02 55.6 & $ -2.637 \pm 0.067 $& $ -27.300 \pm 0.059 $& $ 7.070 \pm 0.034 $& $ 0.9996 $& $ 140.9 ^{+ 2.2 }_{ -2.1 } $& $ -1.8 ^{+ 0.1 }_{ -0.1 } $& $ -18.2 ^{+ 0.3 }_{ -0.3 } $& Off-cloud & \nodata \\
Gaia DR2 6734509617794516480 & 18 27 28.97 & -35  02 58.0 & $ -1.474 \pm 0.061 $& $ -27.993 \pm 0.054 $& $ 6.995 \pm 0.035 $& $ 0.9999 $& $ 142.4 ^{+ 2.2 }_{ -1.9 } $& $ -1.0 ^{+ 0.1 }_{ -0.1 } $& $ -18.9 ^{+ 0.3 }_{ -0.3 } $& Off-cloud & \nodata \\
Gaia DR2 4044800745434225408 & 18 27 52.11 & -34  01 42.1 & $ -1.898 \pm 0.167 $& $ -27.450 \pm 0.145 $& $ 7.086 \pm 0.110 $& $ 0.9999 $& $ 140.6 ^{+ 3.2 }_{ -2.6 } $& $ -1.2 ^{+ 0.1 }_{ -0.2 } $& $ -18.4 ^{+ 0.4 }_{ -0.4 } $& Off-cloud & Class~III \\
Gaia DR2 6734017174039282688 & 18 28 58.78 & -36 51 48.0 & $ -0.795 \pm 0.254 $& $ -28.541 \pm 0.237 $& $ 6.996 \pm 0.144 $& $ 0.9993 $& $ 142.5 ^{+ 3.7 }_{ -3.5 } $& $ -0.5 ^{+ 0.2 }_{ -0.2 } $& $ -19.3 ^{+ 0.5 }_{ -0.6 } $& Off-cloud & Class~III \\
Gaia DR2 4045006594634833152 & 18 29 27.79 & -33  07 47.2 & $ -1.915 \pm 0.299 $& $ -26.697 \pm 0.285 $& $ 6.671 \pm 0.177 $& $ 0.9983 $& $ 149.5 ^{+ 4.9 }_{ -4.2 } $& $ -1.4 ^{+ 0.2 }_{ -0.3 } $& $ -18.9 ^{+ 0.6 }_{ -0.6 } $& Off-cloud & Class~II \\
Gaia DR2 6734929047140217216 & 18 29 27.99 & -34 21 51.1 & $ -1.473 \pm 0.417 $& $ -28.046 \pm 0.375 $& $ 7.197 \pm 0.418 $& $ 0.9983 $& $ 139.3 ^{+ 10.0 }_{ -7.0 } $& $ -0.9 ^{+ 0.3 }_{ -0.4 } $& $ -19.0 ^{+ 1.4 }_{ -1.4 } $& Off-cloud & Transition Disk candidate \\
Gaia DR2 6734227146400420608 & 18 29 34.25 & -35 42 15.5 & $ -0.940 \pm 0.118 $& $ -27.524 \pm 0.106 $& $ 6.701 \pm 0.100 $& $ 1.0000 $& $ 148.7 ^{+ 3.6 }_{ -3.1 } $& $ -0.7 ^{+ 0.1 }_{ -0.1 } $& $ -19.4 ^{+ 0.5 }_{ -0.5 } $& Off-cloud & Class~III \\
Gaia DR2 6734979418518297856 & 18 29 52.07 & -33 55 40.7 & $ -1.540 \pm 0.135 $& $ -28.289 \pm 0.122 $& $ 6.959 \pm 0.082 $& $ 1.0000 $& $ 143.2 ^{+ 2.8 }_{ -2.4 } $& $ -1.1 ^{+ 0.1 }_{ -0.1 } $& $ -19.2 ^{+ 0.4 }_{ -0.4 } $& Off-cloud & Class~III \\

\hline
\hline

\end{tabular}
\tablefoot{For each star, we provide the Gaia-DR2 identifier, position, proper motion and parallax (not corrected for zero-point offset) from the Gaia-DR2 catalogue, membership probability, distance derived from Bayesian inference, 2D tangential velocities in the equatorial system, population, and YSO subclass.}
}
\end{table}
\end{landscape}
\clearpage

\begin{table*}
\centering
\caption{Membership probability for all sources in the field obtained from different probability threshold values for $p_{in}$ used in our analysis. (This table will be available in its entirety in machine-readable form.)
\label{tab_prob}}
\begin{tabular}{cccccc}
\hline\hline
Source Identifier&probability&probability&probability&probability&probability\\
&($p_{in}=0.5$)&($p_{in}=0.6$)&($p_{in}=0.7$)&($p_{in}=0.8$)&($p_{in}=0.9$)\\
\hline\hline
Gaia DR2 6729152036919273472 & 9.3203E-96 & 1.2325E-254 & 5.7880E-136 & 1.8438E-127 & 9.0850E-260 \\
Gaia DR2 6729072842024096128 & 2.2724E-165 & 7.2098E-241 & 3.0367E-203 & 4.7500E-200 & 4.1065E-244 \\
Gaia DR2 6717998590965637632 & 5.5710E-102 & 2.4630E-256 & 1.3444E-143 & 1.3395E-136 & 7.7030E-264 \\
Gaia DR2 6717802774818858112 & 2.7353E-59 & 3.9481E-110 & 1.5472E-73 & 4.4710E-71 & 1.8256E-112 \\
Gaia DR2 6729094248139423360 & 4.7292E-85 & 5.6294E-207 & 1.5074E-112 & 2.6586E-108 & 1.5940E-214 \\
Gaia DR2 6732944806593837824 & 1.7514E-86 & 1.1735E-221 & 7.3986E-122 & 1.5988E-116 & 8.1937E-233 \\
Gaia DR2 6741954823327855104 & 4.4806E-83 & 9.0127E-209 & 1.6230E-113 & 8.9487E-108 & 9.0157E-220 \\
Gaia DR2 6729464336883939712 & 1.4358E-93 & 3.0378E-199 & 8.2241E-123 & 1.9860E-115 & 2.2193E-209 \\
Gaia DR2 6741941045072622848 & 5.5064E-84 & 7.1244E-169 & 6.4287E-108 & 2.7240E-105 & 2.3119E-173 \\
Gaia DR2 6730173998617822208 & 3.6883E-96 & 2.6113E-195 & 5.8357E-124 & 5.6069E-118 & 2.8860E-199 \\
Gaia DR2 6735154893682691840 & 5.0666E-112 & 1.8248E-219 & 7.8541E-147 & 2.8154E-143 & 8.2292E-226 \\
Gaia DR2 4044245281563069568 & 3.8193E-97 & 7.4306E-229 & 3.8310E-133 & 4.6491E-127 & 1.8810E-236 \\
Gaia DR2 6713525021749827456 & 4.8106E-83 & 1.0875E-171 & 3.6448E-109 & 9.0491E-106 & 1.1373E-175 \\
Gaia DR2 6713526675314821760 & 6.8958E-26 & 7.8262E-50 & 7.3366E-33 & 1.0503E-31 & 1.0103E-50 \\
Gaia DR2 6714181017873942784 & 1.7265E-97 & 1.4272E-218 & 2.1920E-132 & 6.7893E-127 & 6.2784E-228 \\
Gaia DR2 6713992833885221376 & 3.0191E-130 & 1.4134E-197 & 7.8871E-155 & 6.9811E-154 & 1.3778E-200 \\
Gaia DR2 6714167273978590464 & 1.1201E-59 & 2.0365E-121 & 1.3166E-77 & 1.4982E-74 & 1.1874E-123 \\
Gaia DR2 6714167484435185664 & 2.3070E-104 & 1.4528E-255 & 6.5021E-143 & 4.4411E-135 & 1.5778E-259 \\
Gaia DR2 6713991317759095424 & 6.0421E-56 & 2.4416E-107 & 3.5321E-71 & 1.3076E-69 & 2.6104E-110 \\
Gaia DR2 6689317628295106816 & 2.4158E-82 & 2.8124E-175 & 4.3123E-106 & 5.3008E-102 & 3.9206E-186 \\
\hline\hline
\end{tabular}
\end{table*}

\begin{table*}
\centering
\caption{Empirical isochrone of the young stars in the Corona-Australis region inferred from our membership analysis. (This table will be available in its entirety in machine-readable form.)
\label{tab_isochrone}}
\begin{tabular}{cc}
\hline\hline
$G_{RP}$&$G-G_{RP}$\\
(mag)&(mag)\\
\hline\hline
4.883	&	-0.328	\\
4.918	&	-0.323	\\
4.953	&	-0.317	\\
4.988	&	-0.312	\\
5.023	&	-0.307	\\
5.058	&	-0.302	\\
5.093	&	-0.297	\\
5.127	&	-0.291	\\
5.162	&	-0.286	\\
5.197	&	-0.281	\\
5.232	&	-0.276	\\
5.267	&	-0.270	\\
5.302	&	-0.265	\\
5.337	&	-0.260	\\
5.372	&	-0.255	\\
5.407	&	-0.250	\\
5.442	&	-0.244	\\
5.477	&	-0.239	\\
5.512	&	-0.234	\\
5.546	&	-0.229	\\ 
\hline\hline
\end{tabular}
\end{table*}

\end{appendix}
\end{document}